\documentclass[twocolumn,english,aps,prl,superscriptaddress]{revtex4-1}
\usepackage[T1]{fontenc}
\usepackage{color}
\usepackage{bm}
\usepackage{amsmath}
\usepackage{amssymb}
\usepackage{graphicx}
\usepackage{array}
\usepackage{mathrsfs}
\usepackage{hyperref}
\makeatletter
\usepackage{bm}
\usepackage{graphicx}

\makeatother

\usepackage{babel}

\begin{document}

\title{Uncover Topology by Quantum Quench Dynamics}

\author{Wei Sun}
\affiliation{Shanghai Branch, National Laboratory for Physical Sciences at Microscale
and Department of Modern Physics, University of Science and Technology
of China, Shanghai 201315, China}
\affiliation{Chinese Academy of Sciences Center for Exellence: Quantum Information and Quantum Physics,
University of Science and Technology of China, Hefei Anhui 230326, China}
\affiliation{CAS-Alibaba Lab for quantum computation, Shanghai 201315, China}

\author{Chang-Rui Yi}
\affiliation{Shanghai Branch, National Laboratory for Physical Sciences at Microscale
and Department of Modern Physics, University of Science and Technology
of China, Shanghai 201315, China}
\affiliation{Chinese Academy of Sciences Center for Exellence: Quantum Information and Quantum Physics,
University of Science and Technology of China, Hefei Anhui 230326, China}
\affiliation{CAS-Alibaba Lab for quantum computation, Shanghai 201315, China}

\author{Bao-Zong Wang}
\affiliation{Shanghai Branch, National Laboratory for Physical Sciences at Microscale
and Department of Modern Physics, University of Science and Technology
of China, Shanghai 201315, China}
\affiliation{Chinese Academy of Sciences Center for Exellence: Quantum Information and Quantum Physics,
University of Science and Technology of China, Hefei Anhui 230326, China}
\affiliation{International Center for Quantum Materials, School of Physics, Peking
University, Beijing 100871, China}
\affiliation{Collaborative Innovation Center of Quantum Matter, Beijing 100871,
China}

\author{Wei-Wei Zhang}
\affiliation{Centre for Engineered Quantum Systems, School of Physics, The University of Sydney, Sydney, NSW 2006, Australia}

\author{Barry C. Sanders}
\affiliation{Shanghai Branch, National Laboratory for Physical Sciences at Microscale
and Department of Modern Physics, University of Science and Technology
of China, Shanghai 201315, China}
\affiliation{Institute for Quantum Science and Technology, University of Calgary, Calgary, Alberta, Canada T2N 1N4}
\affiliation{Program in Quantum Information Science, Canadian Institute for Advanced Research, Toronto, Ontario M5G 1Z8, Canada}

\author{Xiao-Tian Xu}
\affiliation{Shanghai Branch, National Laboratory for Physical Sciences at Microscale
and Department of Modern Physics, University of Science and Technology
of China, Shanghai 201315, China}
\affiliation{Chinese Academy of Sciences Center for Exellence: Quantum Information and Quantum Physics,
University of Science and Technology of China, Hefei Anhui 230326, China}
\affiliation{CAS-Alibaba Lab for quantum computation, Shanghai 201315, China}

\author{Zong-Yao Wang}
\affiliation{Shanghai Branch, National Laboratory for Physical Sciences at Microscale
and Department of Modern Physics, University of Science and Technology
of China, Shanghai 201315, China}
\affiliation{Chinese Academy of Sciences Center for Exellence: Quantum Information and Quantum Physics,
University of Science and Technology of China, Hefei Anhui 230326, China}
\affiliation{CAS-Alibaba Lab for quantum computation, Shanghai 201315, China}

\author{Joerg Schmiedmayer}
\affiliation{Vienna Center for Quantum Science and Technology, Atominstitut, TU Wien, Stadionallee 2, 1020 Vienna, Austria}

\author{Youjin Deng}
\affiliation{Shanghai Branch, National Laboratory for Physical Sciences at Microscale
and Department of Modern Physics, University of Science and Technology
of China, Shanghai 201315, China}
\affiliation{Chinese Academy of Sciences Center for Exellence: Quantum Information and Quantum Physics,
University of Science and Technology of China, Hefei Anhui 230326, China}
\affiliation{CAS-Alibaba Lab for quantum computation, Shanghai 201315, China}

\author{Xiong-Jun Liu}
\affiliation{International Center for Quantum Materials, School of Physics, Peking
University, Beijing 100871, China}
\affiliation{Collaborative Innovation Center of Quantum Matter, Beijing 100871,
China}

\author{Shuai Chen}
\affiliation{Shanghai Branch, National Laboratory for Physical Sciences at Microscale
and Department of Modern Physics, University of Science and Technology
of China, Shanghai 201315, China}
\affiliation{Chinese Academy of Sciences Center for Exellence: Quantum Information and Quantum Physics,
University of Science and Technology of China, Hefei Anhui 230326, China}
\affiliation{CAS-Alibaba Lab for quantum computation, Shanghai 201315, China}

\author{Jian-Wei Pan}
\affiliation{Shanghai Branch, National Laboratory for Physical Sciences at Microscale
and Department of Modern Physics, University of Science and Technology
of China, Shanghai 201315, China}
\affiliation{Chinese Academy of Sciences Center for Exellence: Quantum Information and Quantum Physics,
University of Science and Technology of China, Hefei Anhui 230326, China}
\affiliation{CAS-Alibaba Lab for quantum computation, Shanghai 201315, China}

\begin{abstract}

Topological quantum states are characterized by nonlocal invariants, and their detection is intrinsically challenging. Various strategies have been developed to study topological Hamiltonians through their equilibrium states.  We present a fundamentally new, high-precision dynamical approach, revealing topology through the unitary evolution after a quench from a topological trivial initial state with a two-dimensional Chern band realized in an ultracold $^{87}$Rb atom gas. The emerging ring structure in the spin dynamics uniquely determines the Chern number for the post-quench band and enables probing the full phase diagram of the band topology with high precision. Besides, we also measure the topological band gap and the domain wall structure dynamically formed in the momentum space during the long-term evolution. Our dynamical approach provides a way towards observing a universal bulk-ring correspondence, and has broad applications in exploring topological quantum matter.

\end{abstract}
\maketitle



The discovery of the quantum Hall effect introduced a new fundamental concept, topological quantum phase, to condensed-matter physics~\cite{QHE1980,FQHE1982}. 
In contrast to symmetry-breaking  quantum phases that are characterized by local order parameters in the Landau paradigm,
topological quantum matter is classified by nonlocal topological invariants~\cite{Topo-Order}, which usually do not directly correspond to the local physical observables. In consequence the detection of topological states
is intrinsically challenging. In solid-state experiments, various strategies have been developed
and great success has been achieved in the discovery of topological quantum matter like topological insulators~\cite{topological_insulator, topological_insulator2, QSH2007, DiracQSH2008} and semimetals~\cite{Xia2009, QAHE2013}.
For instance, transport measurements and angle-resolved photoemission spectroscopy are used to
detect gapless boundary modes that reflect the bulk topology~\cite{Weyl1,Weyl2}.
In some circumstances, these strategies do not provide fully unambiguous evidences for topological quantum phases,
as they do not directly measure topological numbers.
An important example is topological superconductivity, which supports a kind of exotic non-Abelian
quasiparticle called Majorana modes~\cite{Majorana1, Majorana2, Majorana3, Majorana4} and remains to be rigorously confirmed by experiment.

To precisely detect the topology for an ultracold-atom system can be more challenging,
whereas the extensive tool box of manipulating and probing ultracold atoms may offer distinct new strategies for measurement. For a one-dimensional (1D) Su-Schrieffer-Heeger model simulated with a 1D double well lattice, the band topology can be determined by measuring the Zak phase~\cite{Zakphase}. Furthermore, the bulk topology of a 2D Chern insulator, characterized by Chern invariants, can be observed by Hall transport studies~\cite{Haldane,Hofstadter}, by Berry curvature mapping~\cite{BerryCur}, and by a minimal measurement strategy~\cite{Liu2013a} of imaging the spin texture at symmetric Bloch momenta~\cite{2DSOC2016}.
The Chern invariants are precisely detectable only in a few special cases, e.g. when the bulk band is flat~\cite{Haldane} or the system is of inversion symmetry~\cite{Liu2013a,2DSOC2016}. 
Nevertheless, the precise detection is typically achieved only for the deep topological regimes. 
Close to the topological critical points, i.e. the topological phase boundaries, 
the measurement becomes increasingly imprecise due to small topological gap and non-ideal conditions like the thermal effects~\cite{2DSOC2017}.

\begin{figure*}[t]
\begin{center}
\includegraphics[width=0.8\linewidth]{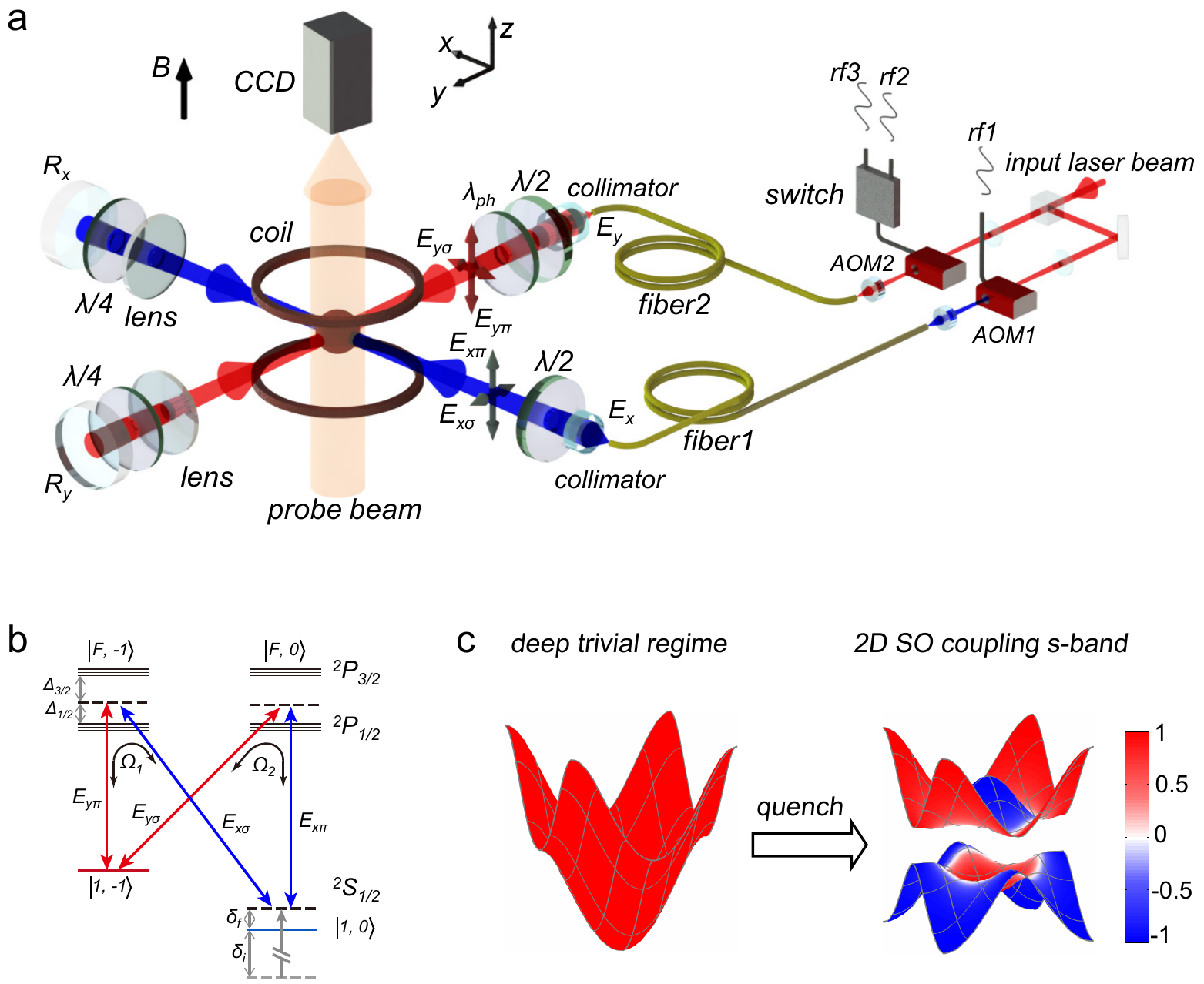}
\vspace{-0.5cm}
\caption{\textbf{Experimental setup and band structure.}
\label{fig_phasediagram}\textbf{a}, Experimental scheme.
The blue and red beams $\boldsymbol{E}_{x}$ and $\boldsymbol{E}_{y}$ incident on the atom cloud represent the lattice lasers.
The probe beam in vertical direction is to detect the momentum distribution of the atoms $($the brown ball$)$ in the $x-y$ plane.
A pair of coils are used to generate the bias magnetic field $\mathbf{B}$ in vertical direction.
All the waveplates ($\lambda/2, \lambda/4$ and $\lambda_{ph}$,) are for realizing the $C_4$ symmetric 2D SO coupling~~({\it28\/}).
The frequency of the lattice beams are controlled by AOM1 and AOM2, with phase locked radio-frequency signal rf1, rf2 and rf3.
The switch is used to perform the quench process by switching rf2 and rf3 to change the two-photon detuning of the Raman couplings.
\textbf{b}, the level structure and the Raman transitions in the experiment.
$\Omega_{1}$ and $\Omega_{2}$ are the two Raman coupling strengths
induced by $($$E_{x\sigma,}E_{y\pi}$$)$ and $($$E_{y\sigma},E_{x\pi}$$)$
respectively. $\delta_{\text{i}}$ and $\delta_{\text{f}}$ are the two-photon detunings
before and after the quenching.
\textbf{c}, The s-band structure before and after the quantum quench. The color represents the distribution of the spin polarization in equilibrium state.}
\end{center}
\end{figure*}

Cold atoms provide realistic platforms to probe quantum dynamics due to the relatively long coherence time. While so far the explorations mostly focus on equilibrium systems, the quench dynamics for the topological systems have been proposed and attracted attention~\cite{quench1,quench2,quench3,quench4,quench5}. In particular, recent experiments~\cite{linking, topoFermion} start to explore the feasibility of observing post-quench band topology by quantum dynamics.
Quenching energy band across topological transition in the Haldane model~\cite{Haldane1988}, a linking number~\cite{quench-topo} is observed in the 3D momentum-time domain characterizing the bulk topology~\cite{linking}. These studies of characterizing band topology of post-quench Hamiltonian are valid for two-band models.

We explore a new dynamical characterization of a 2D quantum anomalous Hall (QAH) insulator.
This lead us to a robust and general strategy to detect the full phase diagram of the band topology.
Our key observation is that if one quenches a 2D QAH system from a trivial to topological regime,
a novel ring pattern in the spin dynamics is observed in momentum space during the unitary evolution,
and uniquely corresponds to the nontrivial bulk topology of the post-quench band, reflecting a dynamical bulk-ring correspondence in Chern insulators~\cite{BulkRing}.
In particular, the different configurations of the pattern, with the ring enclosing $\Gamma$ or $M$ point of the first Brillouin zone (FBZ),
measure the Chern number of the post-quench band.
As a result, we determine the full phase diagram of band topology with a high precision. 
Further, the topological band gap and the dynamical formation of topological domain wall structures in momentum space are also measured to 
obtain the complete information of band topology.

\emph{Quenching the 2D spin-orbit coupled Bose gas}
The present experiment is based on a QAH model~\cite{Liu2014} first realized for a 2D SO coupled gas of ultracold $^{87}$Rb atoms in an optical Raman lattice~\cite{2DSOC2016,Review2018}, with the 2D SO coupling being induced by the periodic Raman potentials along $\hat x$ and $\hat y$ directions (Fig. $1\textbf a$). Note that based on an improved scheme~\cite{Wang2018}, the realized 2D SO coupled quantum gas can have a long life time up to several seconds~\cite{2DSOC2017}, which enable  the present study of quench dynamics. We define the $|1,-1\rangle$ ($|1,0\rangle$) state of $^{87}$Rb as spin up $\left|\uparrow\right\rangle$ (spin down $\left|\downarrow\right\rangle$) (Fig. 1\textbf{b}). The effective Hamiltonian reads
\begin{equation}
	H=\frac{p^{2}}{2m}+V_{{\text {latt}}}(x,y)+\frac{\delta}{2}\sigma_{z}+M_{1}(x,y)\sigma_{x}+M_{2}(x,y)\sigma_{y},
	\label{eq:HSO}
\end{equation}
where $V_{{\text {latt}}}=V_0(\cos^2k_0x+\cos^2k_0y)$ is the potential of the 2D optical lattices with $V_0$ the lattice depth, and $\delta$ is two-photon detuning. $M_{1}=\Omega_0\cos k_0x\sin k_0y$ and $M_{2}=\Omega_0\cos k_0y\sin k_0x$ are the Raman coupling lattices. (For details see Appendix). Without loss of generality we focus on the symmetric case where $M_{1,2}$ are of the same amplitude $\Omega_0$ and the 2D SO coupling is formed with $C_{4}$ symmetry~\cite{Wang2018}.

We commence with a gas of $^{87}$Rb optically pumped into the $\left|\uparrow\right\rangle$ state and cooled to a temperature right above the critical temperature of Bose-Einstein condensation. The atoms are then loaded into a thermal equilibrium states in the 2D optical lattices by adiabatically ramping up the lattice potential in $100\text{ms}$. During this phase the initial two-photon detuning $\delta=\delta_{\text{i}}$ is very large ($\delta_{\text{i}}=-200E_\text{r}$), the Raman coupling lattices $M_1$ and $M_2$ in Eq.~\ref{eq:HSO} are effectively suppressed. The system is fully spin-polarized in a topological trivial band.

\begin{figure*}
\begin{center}
\includegraphics[width=0.8\linewidth]{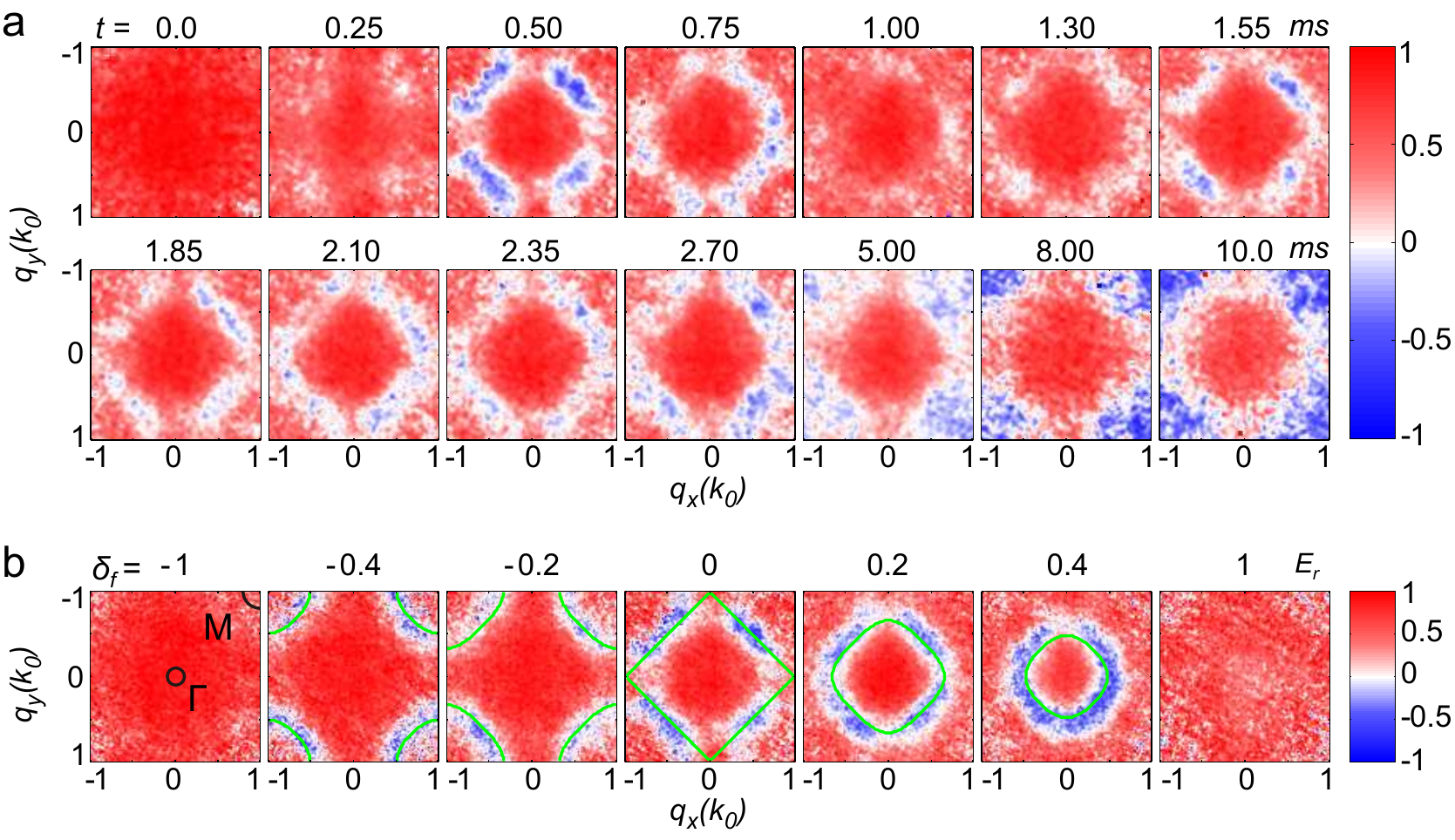}
\vspace{-0.5cm}
\caption{\textbf{Time evolution and \emph{ring} structure of the spin polarization.}  %
\label{Fig_SpinPattern}\textbf{a}. The spin polarization in the first Brillouin zone (FBZ) from $0$ to $10ms$ after quantum quench with the parameters $V_0=4.0E_{\text{r}}$, $\Omega_0=1.0E_{\text{r}}$ and $\delta_{\text{f}}=0E_{\text{r}}$.
\textbf{b}. The same as  \textbf{a} but for fixed time of $480\mu s$ and varying the detuning $\delta_{\text f}$. 
The diagrams with different final two-photon detuning (from left to right):
$\delta_{\text{f}} =-1.0,-0.4,-0.2, 0,0.2, 0.4$ and $1.0E_{\text{r}}$, respectively.}
\end{center}
\end{figure*}

Quench dynamics starts at time $t=0$, when the two-photon detuning $\delta$ for the Raman couplings is switched from its initial value ($\delta_{\text{i}}$) to a final near-resonant value ($\delta_{\text{f}}$) within $200$ns, enables $M_1$ and $M_2$. The band structure of the system changes from the 2D lattice bands to the 2D SO-coupling bands, shown in Fig.~1\textbf{c}.
This rapid quench projects the initial fully spin-polarized states onto superpositions of different eigenstates of the post-quench Hamiltonian $H(\delta_\text{f})$.
This starts a non-equilibrium evolution of Raman-induced Rabi oscillations
between $\left|\uparrow\right\rangle$ and $\left|\downarrow\right\rangle$ states governed by the new Hamiltonian $H(\delta_\text{f})$. The spin oscillation is quantified by the time-evolution of the momentum dependent spin-polarization $P(\bm{q},t)=\frac{N_{\uparrow}-N_{\downarrow}}{N_{\uparrow}+N_{\downarrow}}$, where $\bm q$ is the quasi-momentum, and $N_{\uparrow}$ ($N_{\downarrow}$) denotes the number of atoms at spin up (down) state in quasi-momemtum space .

We measure the spin dynamics after the quench in the following way.
After switching the two-photon detuning from $\delta_{\text{i}}$ to $\delta_{\text{f}}$, we let the system evolve for a certain delay time. 
The spin and momentum distribution of the atomic cloud is obtained by a spin-resolved time-of-flight (TOF) imaging (see Appendix for detail).
Repeating these measurements of the spin-polarization distribution in momentum space for different delay times yields evolution of $P(\bm{q},t)$ at all times and for each momentum.

Before moving onto the experimental observation, we remark on the quench spin dynamics. For convenience we consider the physics in the lowest $s$-band. All the atoms are initialized in the spin-up state. The spin dynamics after the quench is governed by two aspects. First, the Raman coupling potentials $M_{1,2}$, which play the role of a Zeeman field in the $x-y$ plane induce the spin-flip. Second, the local gap $\Delta(\bm {q})$ between spin-up and spin-down bands at quasi-momentum $\bm q$,
defined for the Hamiltonian $H(\delta_{\text f})$ by setting $M_{1,2}\rightarrow0$ (see Appendix for details), serves as a detuning for the dynamics induced by the Raman couplings. For momenta with $\Delta(\bm {q})=0$, the coupling between spin-up and spin-down is resonant, in which case the initial spin-up state at $\bm q$ can fully flip to spin-down in the unitary evolution.


\emph{Uncovering the topology: dynamical ring pattern}
We now turn to analyze this spin dynamics, as observed in the experiment (Fig.~\ref{Fig_SpinPattern}), in different parameter regimes. To get insights from the spin dynamics, we consider first the special case $\delta_{\text{f}}=0$, with the other parameters fixed to $V_0=4.0E_\text{r}$ and $\Omega_0=1.0E_\text{r}$ (Fig. \ref{Fig_SpinPattern}a). The vanishing two-photon detuning $\delta_\text{f}$ implies that the post-quench band at equilibrium is gapless and has two Dirac points at $\{X_{1,2}\}=\{(0,\pi), (\pi,0)\}$. Interestingly, a straight-line pattern of spin-down components connecting the Dirac points emerges in the short-time spin dynamics ($<2$ ms), where the spin evolution is close to unitary as decoherence and dissipation are much slower than the dynamics initiated by the Raman coupling. Such straight-line pattern highlights the momenta where the spin-up and spin-down bands of the Hamiltonian $H(\delta_\text{f}, M_{1,2}\rightarrow0)$ cross, so that $\Delta(\bm q)=0$, and the initial spin-up states can fully flip to spin-down, leading to the observed pattern~\cite{Liu2018}.
After a longer time evolution ($> 10$ ms) the system gradually evolves towards the thermal equilibrium state observed in the previous experiments with adiabatic loading~\cite{2DSOC2016,2DSOC2017}.

When the final two-photon detuning is small but finite, we observe the emergence of a \emph{ring} structure in the spin dynamics. Fig. \ref{Fig_SpinPattern}b shows the spin polarization patterns at a short time ($480\mu\text{s}$) after the quench for a range of $\delta_{\text{f}}$, keeping $V_0=4.0E_\text{r}$ and $\Omega_0=1.0E_\text{r}$ constant. For $\delta_{\text{f}}=0.4E_\text{r}$, spin-polarization $P(\bm q)$ shows a ring pattern circling around the $\Gamma$ point, while for $\delta_{\text{f}}=-0.4E_\text{r}$, $P(\bm q)$ shows \emph{arcs} around the corners of the FBZ, representing a ring circling around the $M$ point. When $\delta_{\text{f}}=0$, the dynamical spin polarization shows of straight lines connecting the two Dirac points ($X_{1,2}$), as discussed above, signifies the transition between the regimes $\delta_\text{f}>0$ and $\delta_\text{f}<0$. In comparison, for large detuning (as $\delta_{\text{f}}=\pm 1.0E_\text{r}$), no dynamical ring pattern appear. Note that different $\delta_{\text{f}}$ leads to the post-quench band exhibiting different topological properties. For $0<|\delta_{\text{f}}|<\delta_\text{c}$, with $\delta_\text{c}$ marks the edge of the topological regime,
the post-quench band is topologically non-trivial with Chern number $\mathcal{C}=\mbox{sgn}(\delta_\text{f})$. $\delta_{\text{f}}=0$ is the boundary separating the topological zones with $\mathcal{C}=\pm1$~\cite{2DSOC2016,Liu2014}. The post-quench band with $|\delta_{\text{f}}|>\delta_c$ is topologically trivial.
These results suggest that the emergence of the ring pattern intrinsically relates to the nontrivial topology of the system, and the ring pattern circling around the $\Gamma$ or $M$ point further gives the sign of the Chern number of the post-quench band. The `broken' \emph{ring} connecting the gapless Dirac points measures the boundary, across which the Chern number varies between $-1$ and $+1$.

\begin{figure*}
\begin{center}
\includegraphics[width=0.8\linewidth]{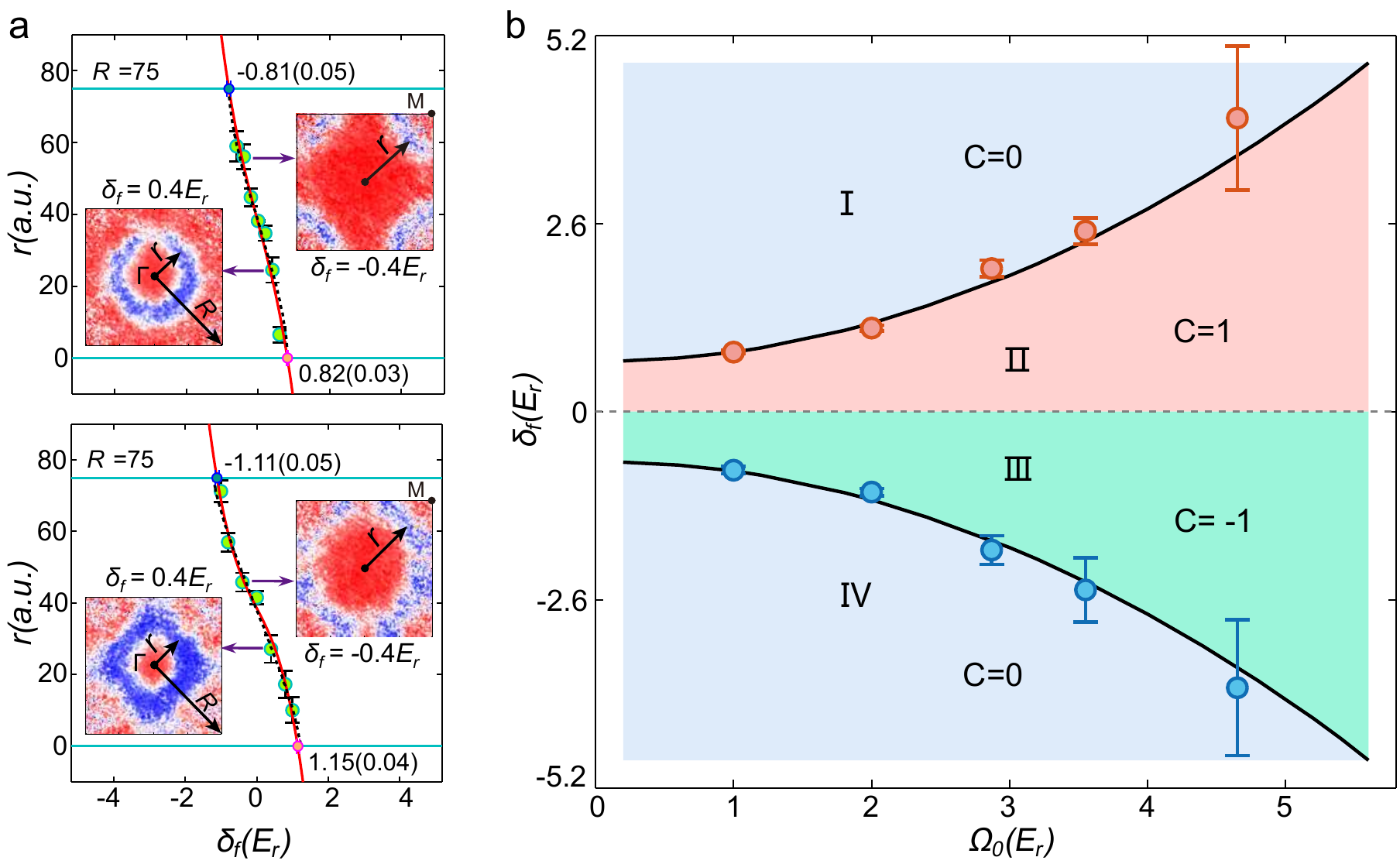}
\vspace{-0.5cm}
\caption{Determination of the topological phase boundary. %
\label{Fig_TopoBoundary}\textbf{a}. Obtaining the topological phase boundaries from the ring structures. The size of the \emph{ring} structure $r$ is measured as shown in the insets. The red line is the polynomial fitting up to 3rd order. The black dot lines are numerical calculation of $r$ versus detuning. The blue lines represent $r = 0$ and $r = R$ and the two cross points (blue dot and red dot) with the red line. The phase boundary is $\delta_{\text c}=0.82\pm0.03E_{\text{r}}$ and $-0.81\pm0.05E_{text{r}}$ for the Raman coupling strength of $\Omega_0= 1.0E_{\text{r}}$ (upper diagram),  and $\delta_{\text c}=1.15\pm0.04E_{{\rm {r}}}$ and $ -1.11\pm0.05E_{\text{r}}$ for $\Omega_0= 2.0E_{\text{r}}$ (lower diagram).
\textbf{b}. Mapping the topological phase diagram in $(\Omega_0, \delta)$ space. The blue and red dots with error-bars are the experimental data determined under different Raman coupling strength $\Omega_0 = 1.0, 2.0, 2.87, 3.55$ and $4.65E_{\text{r}}$. The areas I and IV are topological trivial areas with $\mathcal{C} = 0$ while the areas II and III are topological non-trivial areas with $\mathcal{C} =1$ and $\mathcal{C} = -1$, respectively.}
\end{center}
\end{figure*}

We provide now a theoretical explanation for the above observation. Note that for the SO-coupled Hamiltonian $H(\delta_\text{f})$, the lowest $s$-band physics at equilibrium is captured by a QAH model with inversion symmetry~\cite{Liu2013a,2DSOC2016,Liu2014}. The topology is determined by the product of the spin-polarizations, $\Theta=\prod_{j=1}^{4} \text{sgn}[S(\bf\Lambda_{j})]$, of the lowest $s$-band Bloch states at the four symmetric momenta $\{\bf \Lambda_j\}=\{\Gamma, M, X_{1,2}\}$~\cite{Liu2013a}. Here $S(\bm q)$ denotes the spin texture of the lowest band in equilibrium.
$\Theta=-1$ and $+1$ characterize topological and trivial states, respectively~\cite{Liu2013a}. 
Thus the topological regime necessitates that one of the four spin-polarizations, say $S(\Gamma)$, is opposite to those of the other three. 
Accordingly, the spin-texture $S(\bm q)$ switches sign and must pass through zero when running from the $\Gamma$ point to the other symmetry points. 
All the momenta in the Brillouin zone with $S(\bm q)=0$ form a ring which coincides with the band crossing between the spin-up and spin-down bands defined for Hamiltonian $H(\delta_\text{f}, M_{1,2}\rightarrow0)$. 
Such band crossing is called the band inversion ring~\cite{BulkRing}, across which the spin-up and spin-down bands are inverted~(see also Appendix for detail). 
The band inversion ring corresponds to the vanishing local gap $\Delta(\bm q)=0$ in the dynamical regime, leading to the resonant oscillations between spin-up and spin-down states. 
This uniquely creates the dynamical ring pattern as observed in the experiment. 
The emergence of the dynamical ring pattern therefore is a witness of the nontrivial topology of the bulk band.

Based on the above interpretation, we can precisely determine the Chern number of the post-quench band from the measured dynamics. Note that all atoms are initialized in the spin-up state with large detuning of minus sign. 
The post-quench Hamiltonian $H(\delta_\text{f})$ in the topological regime with a small finite negative value $\delta_\text{f}<0$ ensures that the spin $S(\bf\Lambda_j)$ flips to spin-down for only one of the four symmetric momenta of the lowest band, whereas keeping the other three unchanged.
From Fig.~\ref{Fig_SpinPattern}\textbf{b} we can see that the ring pattern starts to appear circling the $M$ point for the topological regime with $-\delta_\text{c}<\delta_\text{f}<0$, indicating that the spin-polarization of the lowest band at the $M$ point must be negative $S(M)<0$, while the other three keeps positive. Accordingly, the Chern number of the lowest band can be determined directly from $\mathcal C=\bigr[(\Theta-1)/4\bigr]\sum_{j=1}^{4}\text{sgn}[S(\bf\Lambda_{j})]=-1$ in this regime, with $\Theta=-1$~\cite{Liu2013a}. 
Furthermore, if the final detuning $0<\delta_\text{f}<\delta_\text{c}$, the ring pattern encloses the $\Gamma$ point, leading by similar analysis to $S(\Gamma)>0$ and $S(X_{1,2}),S(M)<0$, hence $\mathcal C=+1$. 
With the observation of the ring patterns we can determine the Chern number $\mathcal C=\mbox{sgn}(\delta_f)$ with $0<|\delta_f|<\delta_c$ for the post-quench bands. Note that the spin dynamics is resonant only on the band inversion rings, and the ring patterns can be readily resolved in the experiment.

While the above discussion focuses on spin, which enables an intuitive interpretation for the present inversion-symmetric QAH insulators, characterizing the Chern number ${\cal C}$ by ring patterns is much more general. It can be applied to generic Chern insulators without inversion symmetry and with $|{\cal C}|>1$.
The broad applicability is rooted in the theory~\cite{BulkRing,Liu2018} that the topology of a 2D chiral topological phase characterized by Chern numbers can be uniquely determined by the 1D invariants on the band inversion rings, rendering the bulk-ring correspondence observed in the present experiment. 
As shown in the dynamical classification theory~\cite{BulkRing}, this bulk-ring correspondence is universal 
and the present dynamical approach is applicable to generic multiband systems. 
This is different from the previous dynamical schemes in characterizing the post-quench Hamiltonian by linking number~\cite{linking} and topology-dependent spin relaxation dynamics~\cite{topoFermion}, which are valid for two-band models.

With the relation between dynamical ring and bulk topology being established, we can characterize the full extend of the band topology by observing the size of the ring pattern versus the final two-photon detuning $\delta_\text{f}$. In particular, varying $\delta_\text{f}$ from $0$ to $\delta_\text{c}$, we observe that the size of the ring pattern, as characterized by the distance $r$ between the $\Gamma$ point and the ring [Fig. \ref{Fig_TopoBoundary}\textbf{a}], shrinks and then disappears at the $\Gamma$ point ($r=0$) of $\delta_\text{f}=\delta_\text{c}$, which gives the upper boundary of the topological zone. ($\delta_{\text{f}}=0.82\pm0.03 (1.15\pm0.04)E_{{\rm {r}}}$ for $V_0=4.0E_\text{r}$ at $\Omega_{0}=1.0 (2.0)E_{{\rm {r}}}$, respectively.) On the other hand, varying $\delta_\text{f}$ from $0$ to $-\delta_\text{c}$ increases the distance $r$ until the ring pattern disappears at the $M$ point ($r=R$) for $\delta_\text{f}=-\delta_\text{c}$. This determines the lower boundary of the topological zone, given by $\delta_\text{f}=-0.81\pm0.05 (-1.11\pm0.05)E_{{\rm {r}}}$ at $\Omega_{0}=1.0 (2.0)E_{{\rm {r}}}$. With these measurements we map out the full phase diagram of band topology versus Raman coupling strength, as plotted in Fig. \ref{Fig_TopoBoundary}\textbf{b}. Remarkably, the measured phase diagram is highly consist with the theoretical prediction with the same parameters, showing high precision and the great advantages of the present dynamical approach compared with the previous ones based on equilibrium observing the spin texture~\cite{2DSOC2016,2DSOC2017}.


The great advantages of the present study are rooted in the fundamental difference between characterizations of the topological physics by equilibrium and non-equlibrium. The measurement of topology for static systems necessitates the system to be carefully initialized, with the topological bands being properly occupied, and is therefore sensitive to non-ideal conditions including thermal effects. 
In comparison, for the present dynamical approach the system starts with a fully spin-polarized regime, with the initial spin configuration of Bloch bands being independent of model details. Quenching the system to topological regime leads to the spin dynamics uniquely characterizing the topology of the post-quench bands in short-time (unitary) evolution, which is insensitive to the non-ideal initial conditions.


\begin{figure*}[t]
\begin{center}
\includegraphics[width=0.75\linewidth]{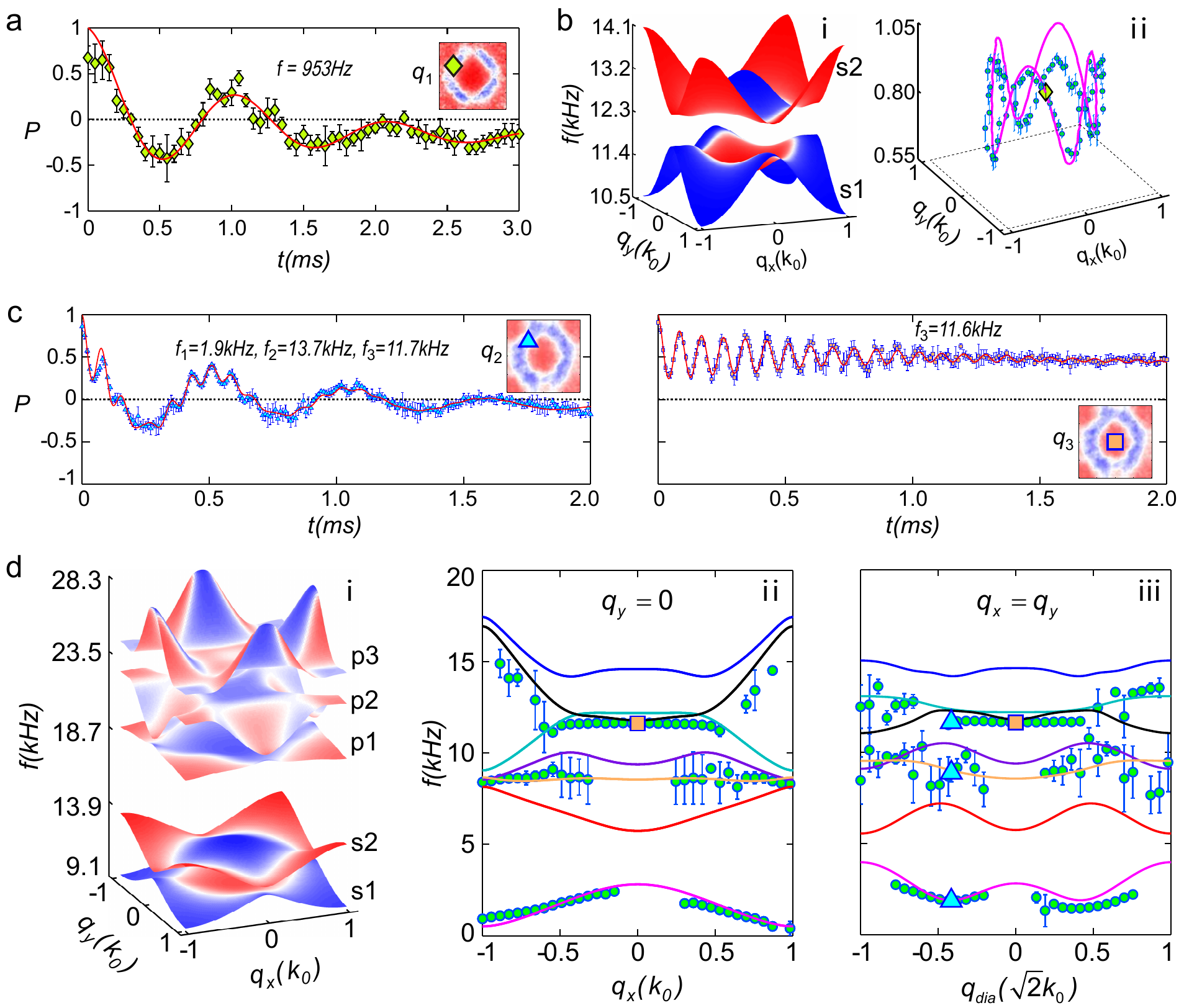}
\vspace{-0.5cm}
\caption{\textbf{Measuring the band structure with quantum quench}
\label{bandmapping}\textbf{a}. Raman-Rabi oscillation at selected \textbf {q} for the parameters: $V_0=4.0E_{\text{r}}, \Omega_0=1.0E_{\text{r}}$ and $\delta_{\text{f}}=0.2E_{\text{r}}$. The insets represent the position of \textbf {q}. The dots with error-bars are experiment data. The red line is the fit.
\textbf{b}. i, the band structure in equilibrium for the same parameter. The color from blue to red represents the spin polarization from $-1$ to $+1$; ii, The dots are experimental data and solid lines are the energy difference between the lowest two bands on the circle of $S(\textbf {q}) = 0$ in the FBZ the green diamond represents the \textbf {q}-point shown in \textbf{a}.
\textbf{c}. Raman-Rabi oscillation of selected $\textbf {q}$-points for the parameters: $V_0=4.0 E_{\text{r}}, \Omega_0=2.0 E_{\text{r}}$ and $\delta_{\text{f}}=0.2 E_{\text{r}}$. The dots with error-bars are experiment data. The red lines are the fit
\textbf{d}. i, Band structure including two s-bands (s1, s2) and three high bands (p1, p2, p3). The color from blue to red represents the spin polarization from $-1$ to $+1$. The band-energy difference between different bands in $q_x$-direction (ii) and diagonal direction (iii): The magenta line is s2-s1, the red line is p1-s2, the yellow line is p1-s1, the purple line is p2-s2, the green line is p2-s1, the black line is p3-s2 and the blue line is p3-s1. The green dots with error-bar are the frequencies obtained from the fitting function. The blue triangle and the orange square represents the \textbf {q}-points shown in \textbf{c}.}
\end{center}
\end{figure*}

\begin{figure*}
\begin{center}
\includegraphics[width=0.82\linewidth]{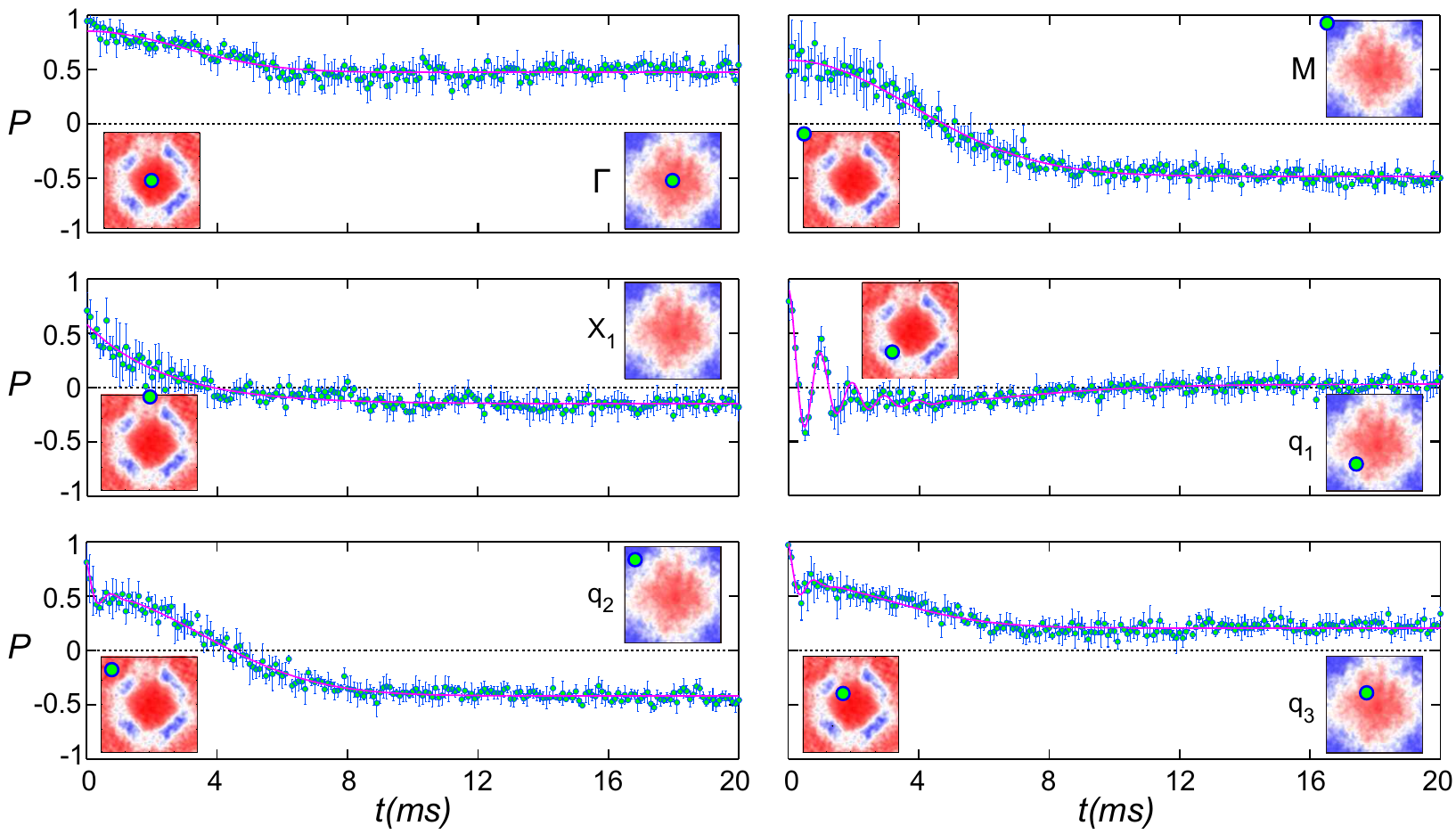}
\vspace{-0.5cm}
\caption{\textbf{Long-term oscillation for different momenta.}  %
\label{Fig_LongTime}The green dots with error bars are the experimental measurement.
The red lines are the fit to the models described in the main text. The measurement is for $V_0=4.0E_{\text{r}}$, $\Omega_0=1.0E_{\text{r}}$ and $\delta_{\text{f}}=0.2E_{\text{r}}$.
The insets are the spin polarizations in the FBZ at $500\mu s$ (left) and $18\text{ms}$ (right) after the quench. 
The spin dynamics exhibit damped oscillation on the ring ($\bm{q}_1$), purely damping behavior at the symmetric points ($\Gamma,M$, and $X_{1,2}$), and weak oscillation in between ($\bm{q}_2$ and $\bm{q}_3$). 
From the long-term evolution a domain wall structure is dynamically formed on the ring. From inside (with $\Gamma$ point) to outside the ring, the long-term spin polarization changes from positive to negative, which corresponds to the Chern number $\mathcal{C}=+1$ for the post-quench band.}

\end{center}
\end{figure*}

\emph{Measuring the topological band gap.}
With the phase diagram of the band topology being mapped out, we discuss now a measurement of the topological band gap in the post-quench band:
The oscillation frequency $\omega(\bm q)$ of the spin dynamics after the quench relates to the energy differences between the populated bands in the post-quench Hamiltonian $H(\delta_\text{f})$~(see Appendix). 
The resonant oscillation occurring on the band inversion ring also corresponds to the scenario of 2D continuous time quantum walk~\cite{ZWW2017}. Measuring $\omega(\bm q)$ throughout the FBZ allows to reconstruct the band structure of the Hamiltonian $H(\delta_\text{f})$, and the band gap corresponds to the minimal oscillation frequency $E_{\rm gap}=\mbox{min}[\omega(\bm q)]$, as studied in Fig.~\ref{bandmapping}.
The experimental data shown in Fig.~\ref{bandmapping}$\textbf a$ is obtained for $V_0=4.0E_{\rm r}$, $\Omega_0=1.0E_{\rm r}$ and $\delta_{\text{f}}=0.2E_{\rm r}$. To extract the distribution of frequencies over the FBZ, we fit the data at every $\bm {q}$ with a decaying oscillating function, as shown with the red curve, which gives the topological band gap $E_{\rm gap}=581\pm61\text{Hz}$. The measurements (blue dots) are compared in Fig.~\ref{bandmapping}$\textbf b$ with the theoretical results (red solid curves) which give the gap $E_{\rm gap}=602\text{Hz}$, with good agreement being confirmed.

The spin polarization dynamics at larger Raman coupling strength $\Omega_0=2.0E_{\rm r}$ apparently exhibits extra high-frequency components besides the original low-frequency one, as shown in Fig.~\ref{bandmapping}$\textbf{c}$. 
The high frequency dynamics is contributed from the transitions between the $s$-band and higher bands. We fit the experimental data with a function consisting of three frequencies and compare them with the energy differences between higher bands along the momentum line $(q_x, q_y=0)$ and $q_x=q_y$, as shown in Fig.~\ref{bandmapping}$\textbf d$. 
From the analysis of the experimental data, the oscillation occurs mainly between ($s1$ and $s2$), ($s1$ and $p1$), ($s2$ and $p3$), respectively. 
In principle, dynamics after the quench allows to reconstruct the SO coupled band structures for both low and higher subbands, which is helpful to investigate the topology of high bands.

\emph {Long-term dynamics to equilibrium.}
Unlike the short-term evolution ($t < 2$ ms) which is nearly unitary, at longer times ($t>2$ ms) a slow relaxation sets in and the spin polarization patterns gradually evolve into an equilibrium distribution, as shown in Fig.~\ref{Fig_LongTime}.

For the $\Gamma$, $M$ and $X_1$ points, the Raman coupling strength vanishes~(see Appendix), while the relatively large $\Delta(\bm {q})$ dominates the spin dynamics. The spin polarizations at these points experience no oscillations but a slow relaxation with Gaussian ($\Gamma$ and $M$) or exponential ($X_{1}$) shape from the initial spin-up state to their final values.
For quasi-momentum $\bm{q}_{1}$ on the \emph{ring}, the local gap $\Delta (\bm{q}_{1})=0$ and the Raman coupling dominates the spin oscillation. 
The spin polarization experiences a damped oscillation with large amplitude followed by a slow relaxation towards the equilibrium spin polarization of $P(\bm{q}_1,t \rightarrow \infty)=0$. 
This is consistent with the dynamical characterization of band inversion ring~\cite{BulkRing}. 
Out of the ring ($\bm{q}_{2,3}$ in Fig.~\ref{Fig_LongTime}), 
the spin polarization approaches nonzero values in the long-term evolution and changes sign across the ring, implying a domain wall structure dynamically formed in the momentum space (insets of the long term spin-polarization in the FBZ in Fig.~\ref{Fig_LongTime}). The formed domain wall structure corresponds to the bulk topology with Chern number $\mathcal C_1=+1$ for $\delta_f=0.2E_{\text{r}}$ in the post-quench band.

The long-term relaxation is governed mainly by the external field induced dephasing and the atom-atom interaction induced decay.
The dissipation in the oscillation at $\bm{q}_1$ mainly resulted from the noise of the bias magnetic field of typical $\delta B_{\rm rms}\approx140\mu\text{G}$, which causes an uncertainty in the Raman detuning to about $100\text{Hz}$ and a related de-phasing in the order of $1/(2 \pi\times 100 \text{Hz}) \sim 1.6$ms. The slower relaxation with a Gaussian shape, mostly visible outside the \emph {ring} ($\Gamma$, $M$, and $\bm{q}_{2,3}$ points), is attributed to the interaction between the atoms~\cite{Decaymodel}. 
The interaction strength can be achieved from the time scale of the relaxation.
In our present experiment, the relaxation time is about $5.3$ms, correspond to the interaction energy per atom to be about $30\text{Hz}$ (see Appendix for detail). It is consist with the atom density in our system, which is about $4\times 10^{12} \text{cm}^{-3}$.
The long-term relaxation leads to `thermalized' atom and spin distribution, similar to the state reached by 
the adiabatic loading used in the previous experiments~\cite{2DSOC2016,2DSOC2017} with extra heating. 
It is also possible to extend our approach for studying the topological phases of a correlated or interacting system.


\emph{Conclusion and Discussions.}
The present dynamical measurement of topological quantum physics exhibits clear advantages over
the previous measurements using static phases.
Measurement of topology for static systems is by nature highly dependent
on the initialization of the equilibrium system~\cite{2DSOC2016,2DSOC2017},
and is therefore sensitive to imperfect conditions including thermal effects which may ruin the precise detection of the bulk topology.
In comparison, for the dynamical measurement the system can be prepared in a fully spin-polarized trivial regime,
with the initial spin distribution of Bloch bands being independent of the details.
Quenching the system to a topological regime leads to spin dynamics which is governed
by the post-quench topological Hamiltonian, with the short-term unitary dynamics not sensitive to non-ideal conditions.
This enables a high-precision measurement of the full phase diagram of band topology by observing the emergence
 and configuration of ring patterns of the spin dynamics.

The observed {\it ring} patterns are the evidence of 
the so-called {\it bulk-ring correspondence} introduced in Ref.~\cite{BulkRing},
a fundamental and universal concept that can be applied in generic QAH insulators with high Chern numbers and multibands.
Thus, the present quench-dynamic approach is intrinsically different from the previous strategy~\cite{2DSOC2016, Liu2014}
which measures Chern bands from spin texture at symmetric momenta
and is only valid for systems with low Chern invariants $C=0,\pm1$~\cite{Liu2013a}.
Further, the present dynamical approach is also fundamentally different from the linking number characterization that is valid in two-band models~\cite{linking,quench-topo}.
More generically, the present study reveals an important insight that a higher dimensional topological system (2D Chern insulator)
can be characterized by a lower dimensional invariant (1D ring pattern) on the band inversion ring.
This insight can be extended to generic topological phases characterized by $Z$ invariants of any dimension,
and gives a deep dynamical classification theory of the broad classes of topological quantum states~\cite{BulkRing}. 
In particular, for the 1D SO coupled topological phase, the bulk topology can be readily determined by observing the dynamical spin pattern at the two band inversion points. More interestingly, our approach can be applied to detecting 3D topological phases, e.g. the 3D topological semimetal phases. 
Note that the band topology of a 3D system is hard to detect, since the conventional mentum-space tomography method or band mapping technique is valid only to map out 1D/2D band structures. 
With the present approach we can reduce the band topology of a 3D system to the lower-dimensional band inversion surfaces, which should be measurable and may solve the essential difficulty in mapping out the topology for high dimensional states~\cite{BulkRing}.
The present dynamical approach should also be applicable to interacting topological phases whose topology can be characterized by quasiparticles or mean field picture, so that the band inversion surfaces/rings are still well defined.
Therefore, our work establishes an insightful and powerful approach to explore novel topological quantum states with non-equilibrium dynamics.

\emph{Acknowledgement.}
We acknowledge insightful discussions with Ting-Fung Jeffrey Poon, Hui Zhai, and Jinlong Yu.
This work was supported by the National Key R\&D Program of China (under grants 2016YFA0301601 and 2016YFA0301604), National Nature Science Foundation of China (under grants N0. 11674301, No. 11761161003, and No. 11625522), and the Thousand-Young-Talent Program of China.
WWZ is supported by the Australian Research Council (ARC) via the Centre of Excellence in Engineered Quantum Systems (EQuS) project number CE110001013. BCS acknowledges China's 1000 Talent Plan and NSFC (Grant No. GG2340000241) for support. JS acknowledges support by the European Research Council, ERC-AdG {\em{QuantumRelax}}.



\newpage

\onecolumngrid

\newpage{}

\section*{\large Appendix}

In this Appendix we provide the details of experimental realization on 2D SO coupling, the quench process with detecting method, the general theory on the relation between the topology and the \emph{ring}, the theory on the quantum quench as a continuous-time quantum walk, the details on the short time evolution, and analysis the long-term evolution with decay.

\setcounter{equation}{0} \setcounter{figure}{0} \setcounter{table}{0}
\setcounter{page}{1} \makeatletter \global\long\def\theequation{S\arabic{equation}}
 \global\long\def\thefigure{S\arabic{figure}}

\section*{I. Experimental Procedure}

\subsection*{A. Generation of a 2D Spin-Orbit coupled system for Bose gas}

The generation of a 2D SO coupled Bose gas is realised with the $^{87}$Rb atoms trapped and
cooled in an crossed-beam optical dipole trap as shown in Fig. 1\textbf{a}~\cite{2DSOC2017,Wang2018}.
The spin up and down basis are defined by two magnetic sub-levels in $F=1$ manifold, i.e.~$\left|\uparrow\right\rangle\equiv\left|1,-1\right\rangle$ and $\left|\downarrow\right\rangle\equiv\left|1,0\right\rangle$, which are resulted from the $10.2{\rm {MHz}}$ Zeeman splitting generated with a bias magnetic field ($\boldsymbol{B}$) of $14{\rm {G}}$ applied along the $\hat{z}$ direction. The large quadratic Zeeman shift of $\epsilon\approx8E_{{\rm {r}}}$ makes the coupling to the $|1,1\rangle$ state negligible in our experiment. Then it can be treated as an effective two level (spin 1/2) system.

The 2D square lattice $V_{{\rm {latt}}}(x,y)$ is constructed with two laser beams $\boldsymbol{E}_{x}$
and $\boldsymbol{E}_{y}$ noted as the blue and red lines in Fig.~1\textbf{a}.
The wavelength of the beams is around $\lambda=787\text{nm}$ and the frequency difference is close to $10.2\text{MHz}$.
The laser beams $\boldsymbol{E}_{x}$
and $\boldsymbol{E}_{y}$  injected from $\hat{x}$ and $\hat{y}$ directions pass through a high extinction ratio PBS (polarized beam splitter) separately and are further rotated by a $\lambda/2$ wave plate to obtain the linear
polarization component $E_{x\pi}$ ($E_{y\pi}$) and circular component $E_{x\sigma}$ ($E_{y\sigma}$), respectively.
The two beams retro-reflected back by two mirrors ($R_x$ and $R_y$) to form a 2-dimensional square lattice.
Two $\lambda/4$ wave plates with optical axis along $\hat{z}$ direction are placed in front of the two mirrors as phase retarders to generate $\pi/2$ phase delay between the two lattices with two orthogonal polarization components.
Then, the 2D lattice have the form of
\begin{equation}
V_{{\rm {latt}}}(x,y)=V_{0x}\cos^{2}k_{0}x+V_{0y}\cos^{2}k_{0}y,
\end{equation}
with $V_{0x}\propto |E_{x\pi}|^{2}-|E_{x\sigma}|^{2}$ and $V_{0y}\propto|E_{y\pi}|^{2}-|E_{y\sigma}|^{2}$~\cite{2DSOC2017,Wang2018}.

The orthogonal polarization components, ($E_{x\pi}, E_{y\sigma}$) and ($E_{x\sigma}, E_{y\pi}$), generate  two independent Raman couplings lattices $M_1=\Omega_1\sin{k_{0}x}\cos{k_{0}y}$ and $M_2=\Omega_2\cos{k_{0}x}\sin{k_{0}y}$.
An additional wave plate $\lambda_{\text{ph}}$ is inserted in one of the beam paths to generate the relative phase $\delta \varphi_\text{L}$ between the $M_1$ and $M_2$ terms, which gives the Raman coupling term
\begin{equation}
V_\text{R}(x,y)=(M_1-M_2\cos{\delta\varphi_\text{L}})\sigma_x+M_2\sin{\delta\varphi_\text{L}}\sigma_y.
\end{equation}
The two-photon detuning $\delta$, which is precisely controlled with two phase-locked acousto-optical modulators (AOM1 and AOM2), generates the detuning term $\frac{\delta}{2}\sigma_z$.
In our experiment, we tune $V_{0x}=V_{0y}=V_0$, $\Omega_1=\Omega_2=\Omega_0$ and $\delta\varphi_L=\pi/2$; thus the Hamiltonian has a pure SO coupling having $C_4$ symmetry as shown in Eq. (1), with the band structure shown in Fig.~1\textbf{c}.

Our setup realizes the highly controllable and robust 2D SO coupling Bose gas~\cite{2DSOC2017}, with life time up to several seconds.
With $C_4$ symmetry intrinsically satisfied, our system explores the deep topological nontrivial phase with large Raman coupling.
The lattice laser path is set up in a very compact way, which suppresses the mechanical vibration of the laser used to manipulate the atoms.
Moreover, the magnetic noise is well shielded with a $\mu$-metal and  further cancelled with real-time feedback technique, which reduces the heating on atoms and enhances the lifetime of atoms to several seconds.


\subsection*{B. Quantum quench dynamics with $^{87}\text{Rb}$ in a 2D SO system}

About $1\times10^{5}$ of $^{87}$Rb atoms at $100{\rm {nK}}$ prepared in the spin-up state $\left|\uparrow\right\rangle\equiv\left|1,-1\right\rangle$ are adiabatically loaded into the 2D square-lattice potential $V_{{\rm {latt}}}(x,y)$ by ramping up the laser intensity (about 100ms), where the initial two-photon detuning $\delta_\text{i}=-200E_\text{r}$ is large enough to exclude the Raman coupling.
The atoms fill the normal lattice potential and stay in the thermal equilibrium state of the lattice Hamiltonian. This is the process of initial-state preparation.

To initialise the quench we switch at time $t=0$ the control RF signal of the AOM2 from rf2 to rf3 (Fig.~1\textbf{a}), which changes the two-photon detuning $\delta$ from $\delta_\text{i}$ to near resonance $\delta_\text{f}$ in 200ns, and the two-photon Raman coupling is turned on simultaneously. This is the starting point for a non-equilibrium dynamics.
The evolution is close to unitary as decoherence and dissipation are much slower than the dynamics initiated by the Raman coupling in the beginning.
The band structure of the system changes at $t=0$ from a pure lattice bands to 2D SOC bands, which shown in Fig.~1\textbf{c}.
In quench dynamics, the atoms in $\left|\uparrow\right\rangle$ state are driven by Raman coupling and start the Raman-Rabi oscillation in the 2D SO coupled system's bands; i.e. the internal state oscillation between $\left|\uparrow\right\rangle$ and $\left|\downarrow\right\rangle$.

\subsection*{C. Spin-resolved momentum detection}

\begin{figure}[h]
\includegraphics[width=0.75\linewidth]{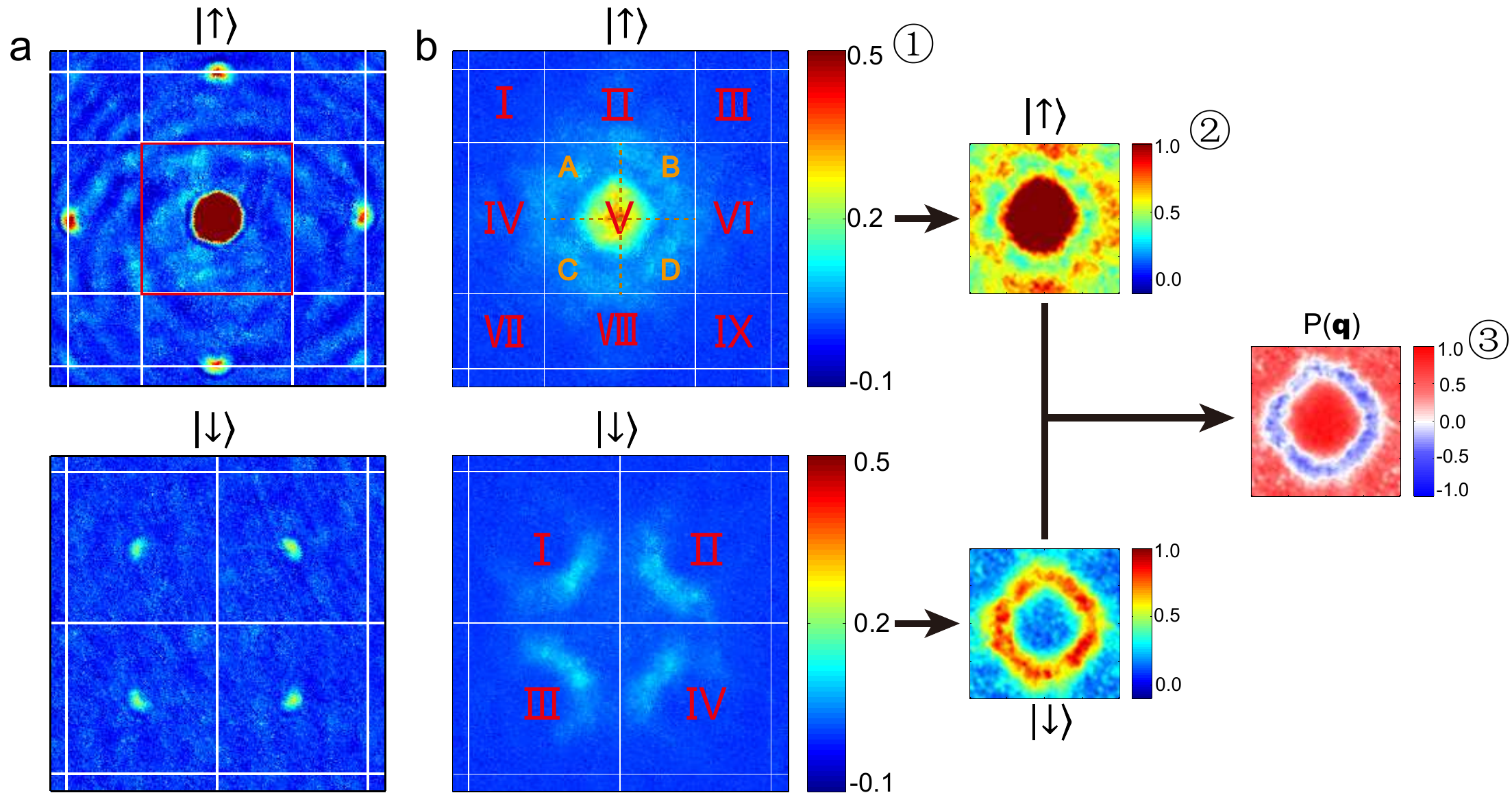}
\caption{\textbf{Spin-resolved momentum detection.}
 \textbf{a}. Image of a BEC in the lattice after TOF. The upper image shows the mechanical momentum distribution of the spin-up atoms. The center atoms cloud has zero mechanical momentum. The four diffracted atoms clouds around are the first order lattice diffractions. The red rectangle represents the size of the FBZ. The white lines show the  distribution of the FBZ in the mechanical momentum detection. The lower image shows the mechanical distribution of the spin-down atoms, which are the first Raman diffractions. \textbf{b}. The steps to obtain the spin polarization in the FBZ. step\textcircled{1}: The white lines cut the raw images into nine regions for spin-up and four parts for spin-down atoms. step\textcircled{2}: The density distributions of spin-up and spin-down atoms in the FBZ. step\textcircled{3}: The spin polarization in the FBZ obtained by substracting and normalizing the density distributions of spin-up and spin-down atoms.}
\label{figureFBZ}
\end{figure}

The observation of the quench dynamics is achieved by spin-resolved time-of-flight (TOF) absorption imageing. The dipole trap and lasers for 2D lattices and Raman coupling are all simultaneously switched off  within $1{\rm {\mu s}}$, and the atoms are projected back to the bare states $\left|1,-1\right\rangle$ and $\left|1,0\right\rangle$ while keeping the same momentum distribution in the lattices.
The TOF absorption image is taken after  25 ms of free expansion in a Stern-Gerlach gradient magnetic field applied along the $x$-axis to obtain the separation of $\left|1,-1\right\rangle$ and $\left|1,0\right\rangle$ in the horizontal direction. The image resolves the spin and the momentum distribution of the atoms simultaneously as shown in Fig.~\ref{figureFBZ}.

While the TOF image shows the mechanical momentum distribution, we need to transfer it to the FBZ.
The eigen-state is a Bloch function which has the form of
\begin{equation}
|\Psi\rangle=|\Psi^{\uparrow}\rangle+|\Psi^{\downarrow}\rangle=\sum_{m,n}a_{m,n}\psi^{\uparrow}_{m,n}\chi^{\uparrow}
+\sum_{p,q}b_{p,q}\psi^{\downarrow}_{p,q}\chi^{\downarrow}.
\end{equation}
In the Bloch function, $|\Psi^{\uparrow}\rangle=\sum_{m,n}a_{m,n}\psi^{\uparrow}_{m,n}\chi^{\uparrow}$ represents the state for spin-up atoms and $|\Psi^{\downarrow}\rangle=\sum_{p,q}b_{p,q}\psi^{\downarrow}_{p,q}\chi^{\downarrow}$ represents the state for spin-down atoms, where
\begin{equation}
\psi^{\uparrow}_{m,n}=\frac{1}{\sqrt{S}}e^{i(q_{x}+\frac{2\pi}{a}m)x}e^{i(q_{y}+\frac{2\pi}{a}n)y}
\end{equation}
and
\begin{equation}
\psi^{\downarrow}_{p,q}=\frac{1}{\sqrt{S}}e^{i(q_{x}+\frac{2\pi}{a}p+\frac{\pi}{a})x}e^{i(q_{y}\
+\frac{2\pi}{a}q+\frac{\pi}{a})y}.
\end{equation}
Here $a$ is the lattice constant and $\frac{\pi}{a}$ equals to the wave vector $k_{0}$ and the FBZ belongs to $(q_{x}\in [-k_{0}, k_{0}]$, $q_{y}\in [-k_{0}, k_{0}])$. It can be found that the eigen-state at qausi-momentum $(q_{x}, q_{y})$ is the superposition of $(q_{x}+2mk_{0}, q_{y}+2nk_{0})$-states for spin-up atoms and $(q_{x}+(2p+1)k_{0}, q_{y}+(2q+1)k_{0})$-states for spin-down atoms. The extra momentum $k_{0}$ shift for spin-down atoms comes from the Raman coupling, which transfers the photon momentum to the atoms.

In Fig.~\ref{figureFBZ}\textbf{a}, a BEC is in the ground state of the lattice at the $\Gamma$ point with quasi-momentum $(q_{x}=0, q_{y}=0)$. After TOF, the spin-up atoms show odd peaks of $(2mk_{0}, 2nk_{0})$ which are the $(m,n)$-order lattice diffractions. The spin-down atoms show the Raman diffractions with momentum of $((2p+1)k_{0}, (2q+1)k_{0})$ . We determine the size of the FBZ from the atom diffraction after TOF as the red rectangle. In Fig.~\ref{figureFBZ}\textbf{b}, thermal atoms fill the whole FBZ in quasi-momentum space. After TOF, the atoms distribute broadly. To map the atoms into the FBZ, we firstly cut the spin-up image into the array of $([-k_{0}, k_{0}]+2mk_{0},[-k_{0}, k_{0}]+2nk_{0})$ and the spin-down image into the array of $([-k_{0}, k_{0}]+(2p+1)k_{0},[-k_{0}, k_{0}]+(2q+1)k_{0})$. Then all the different regions in the array are mapped back into the FBZ, respectively. In step \textcircled{1}, for the spin-up thermal atoms, the raw image is cut into nine regions: I to IX. For the spin-down thermal atoms, the raw image is cut into four regions: I to IV. In step \textcircled{2}, the nine closets for spin-up atoms are moved into the FBZ in the way: I$\to$D, II$\to$CD, III$\to$C, IV$\to$BD, V unchanged, VI$\to$AC, VII$\to$B, VIII$\to$AB, IX$\to$A. And the four closets for spin-down atoms are all moved to ABCD. After moving each closet, the atomic distributions in each closet are summed together to get the density distributions of spin-up and spin-down atoms in the FBZ, respectively. In step \textcircled{3}, we directly substract the density distributions of spin-up and spin-down atoms at every momentum in the FBZ and make the normalization. Then we get the distribution of the spin polarization $P(\bm{q}, t)=\frac{N_{\uparrow}-N_{\downarrow}}{N_{\uparrow}+N_{\downarrow}}$.


\section*{II. Theory}

\subsection*{A. The generic formalism: Determining Chern number from the ring}

\begin{figure}[h]
\includegraphics[scale=0.8]{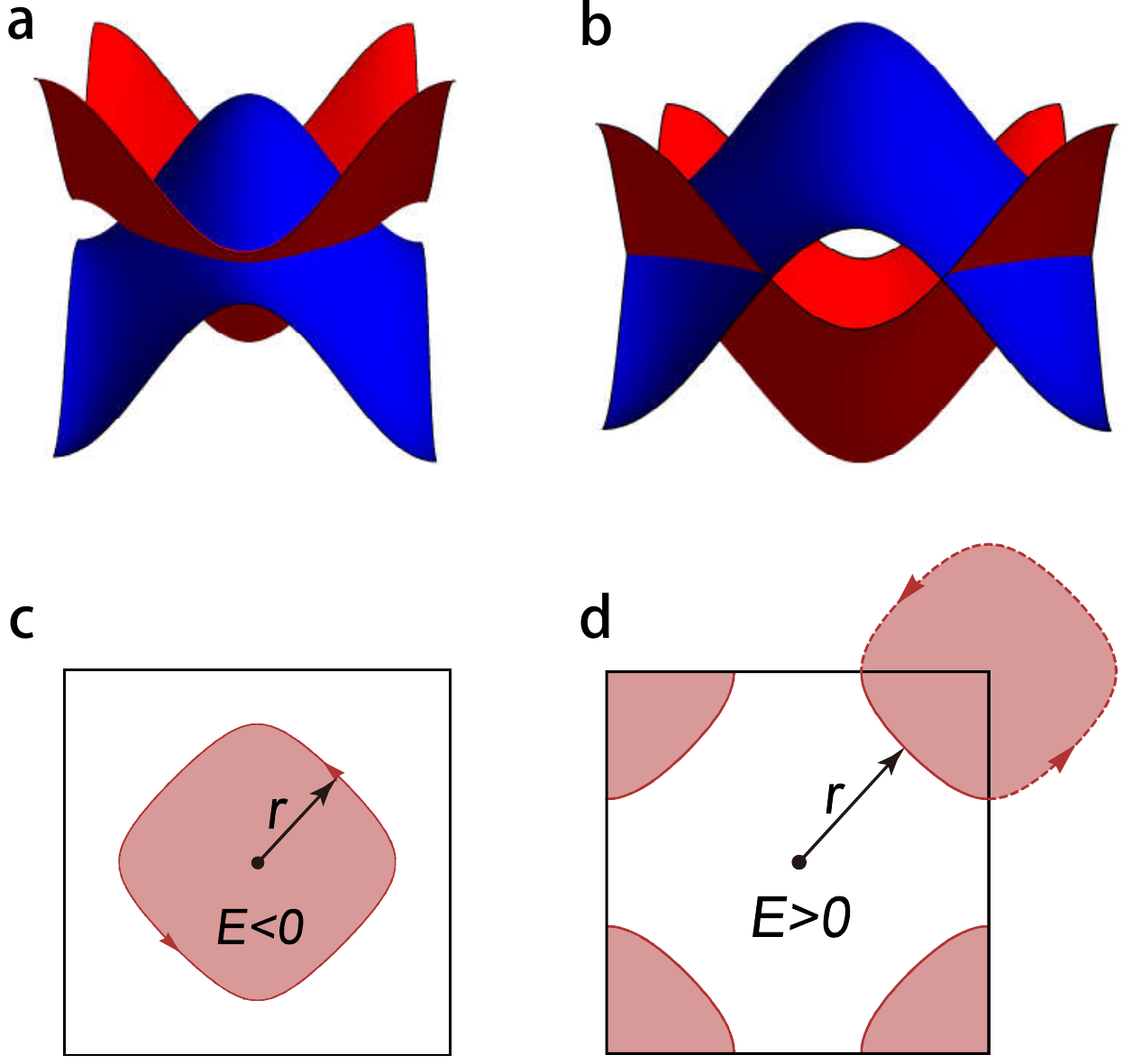}
\caption{\textbf{The normal bands in our system.} \textbf{a}. The normal bands for the case
$\delta_{\text{f}}>0$, and the corresponding ring \textbf{c} in the First Brillouin
Zone. \textbf{b}. The normal bands for the case $\delta_{\text{f}}<0$, and the corresponding
ring \textbf{d} in the Brillouin Zone. }
\label{bandtheory}
\end{figure}

Here we show a generic theorem to determine the Chern number of the topological
band based on the observation of the ring. This theorem was first proven in the
paper~\cite{Liu2018} for chiral superfluid phases characterized by Chern numbers, and can be directly applied to the present insulating phases with minor modification. We first consider the two-band system, with the Hamiltonian
being generically written as
\begin{alignat}{1}
\mathcal{H}(\bm{q}) & =\begin{pmatrix}\epsilon_{\bm{q}\uparrow} & M_{\bm{q}}\\
M_{\bm{q}}^{*} & \epsilon_{\bm{q}\downarrow}
\end{pmatrix},
\end{alignat}
where $\epsilon_{\bm{q}\uparrow}$ and $\epsilon_{\bm{q}\downarrow}$
represent the normal bands (without the coupling), and the off-diagonal
term $M_{\bm{q}}$ denotes the coupling between two normal bands (mimicing the pairing terms in the chiral topological superfluid phases~\cite{Liu2018}).
For the lowest band, the energy and the corresponding eigen state
are
\begin{align}
\epsilon_{-}(\bm{q}) & =\frac{1}{2}\left(\epsilon_{\bm{q}\uparrow}+\epsilon_{\bm{q}\downarrow}-\sqrt{(\epsilon_{\bm{q}\uparrow}-\epsilon_{\bm{q}\downarrow})^{2}+4|M_{\bm{q}}|^{2}}\right)\\
\psi_{-}(\bm{q}) & =\frac{1}{N(\bm{q})}\begin{pmatrix}\epsilon_{\bm{q}\uparrow}-\epsilon_{\bm{q}\downarrow}-\sqrt{(\epsilon_{\bm{q}\uparrow}-\epsilon_{\bm{q}\downarrow})^{2}+4|M_{\bm{q}}|^{2}}\\
2M_{\bm{q}}^{*}
\end{pmatrix},
\end{align}
where $N(\bm{q})$ is a normalization factor. To obtain the Chern
number of the lowest band, we need to calculate the Berry connection
\begin{align*}
\mathcal{A}_{\bm{q}} & =i\psi_{-}^{*}(\bm{q})\nabla_{\bm{q}}\psi_{-}(\bm{q})\\
 & =\frac{M_{\bm{q}}\nabla_{\bm{q}}M_{\bm{q}}^{*}-M_{\bm{q}}^{*}\nabla_{\bm{q}}M_{\bm{q}}}{\sqrt{(\epsilon_{\bm{q}\uparrow}-\epsilon_{\bm{q}\downarrow})^{2}+4|M_{\bm{q}}|^{2}}\left(\sqrt{(\epsilon_{\bm{q}\uparrow}-\epsilon_{\bm{q}\downarrow})^{2}+4|M_{\bm{q}}|^{2}}+\epsilon_{\bm{q}\downarrow}-\epsilon_{\bm{q}\uparrow}\right)}.
\end{align*}
The topology of the band cannot be changed by a continuous deformation without closing a gap.
Hence, we can deform $M_{\bm{q}}\rightarrow\gamma M_{\bm{q}}$ and let
$\gamma\to0$, in which limit the bands cross, i.e. the ring appears at $\epsilon_{\bm{q}\uparrow}=\epsilon_{\bm{q}\downarrow}$. The Berry connection in this limit is given by
\begin{align}
\mathcal{A}_{\bm{q}} & =\frac{i\Theta_{\mathcal{ S}}}{2}\frac{M_{\bm{q}}\nabla_{\bm{q}}M_{\bm{q}}^{*}-M_{\bm{q}}^{*}\nabla_{\bm{q}}M_{\bm{q}}}{|M_{\bm{q}}|^{2}},
\end{align}
where $\mathcal{S}$ is a vector area enclosed by the ring, $\Theta_{\mathcal{ S}}$
denotes the step function that is $1$ inside (outside) the region $\mathcal{S}$
and $0$ outside (inside)  $\mathcal{S}$ if the spectra
within the area $\mathcal{S}$ satisfies $\epsilon_{\bm{q}\uparrow}>\epsilon_{\bm{q}\downarrow}$ ($\epsilon_{\bm{q}\uparrow}<\epsilon_{\bm{q}\downarrow}$). The Berry curvature
reads $\mathcal{B}_{\bm{q}}=\bm{\nabla}_{\bm{q}}\times\mathcal{A}_{\bm{q}}$
and the Chern number can be calculated by the integration of $\mathcal{B}_{k}$ that
\begin{align*}
\mathcal{C} & =\frac{1}{2\pi}\int\mathcal{B}_{\bm{q}}{\rm d}^{2}\bm{q}=\frac{1}{2\pi}\oint\mathcal{A}_{\bm{q}}\cdot{\rm d}\bm{q}\\
 & =(-1)^{m}\frac{1}{2\pi}\text{\ensuremath{\oint}}_{\partial\mathcal{S}}\nabla_{\bm{q}}\theta_{\bm{q}}\cdot{\rm d}\bm{q}\\
 & =(-1)^{m}n.
\end{align*}
where $\partial\mathcal{S}$ represent the ring, which is also the
boundary of the region $\mathcal{S}$. The $\theta_{\bm{q}}={\rm arg}(M_{\bm{q}})$
is the phase of Ramman coupling projected onto the ring and $n$ is
the winding number of $\theta_{\bm{q}}$, here we always take this integration
along a anticlockwise direction as shown in Fig.~\ref{bandtheory}, and the quantity
$m=0$ $(\text{or }1)$ if the spectra satisfy $\epsilon_{\bm{q}\uparrow}<\epsilon_{\bm{q}\downarrow}$ inside the
ring (or $\epsilon_{\bm{q}\uparrow}>\epsilon_{\bm{q}\downarrow}$ inside the ring).

The above result can be generalized to the case with multiple rings, which is the same as the case for chiral superfluids with multiple Fermi surfaces~\cite{Liu2018}, giving the
total Chern number
\begin{align*}
\mathcal{C} & =\frac{1}{2\pi}\int\mathcal{B}_{\bm{q}}{\rm d}^{2}\bm{q}\\
 & =\sum_{i}(-1)^{m_{i}}\frac{1}{2\pi}\oint_{i}\nabla_{\bm{q}}\theta_{\bm{q}}^{i}\cdot{\rm d}\bm{q}\\
 & =\sum_i (-1)^{m_{i}}n_{i}.
\end{align*}
where $i$ denotes the $i$-th ring. The quantity $m_{i}=0$ or $1$ depending on the condition presented above. The $\theta_{\bm{q}}^{i}={\rm arg}(M_{\bm{q}})$ is the
phase of Ramman coupling projected onto the $i$-th ring. Hence, the
total Chern number is just the summation the winding numbers with proper sign factors.

Finally, for the case of more than one band intersecting with the lowest band,
the conclusion is still the same. The detailed proof can be found
in the paper~\cite{Liu2018}.

\subsection*{B. Applying to the present system}

\begin{figure}[h]
\includegraphics[width=0.9\linewidth]{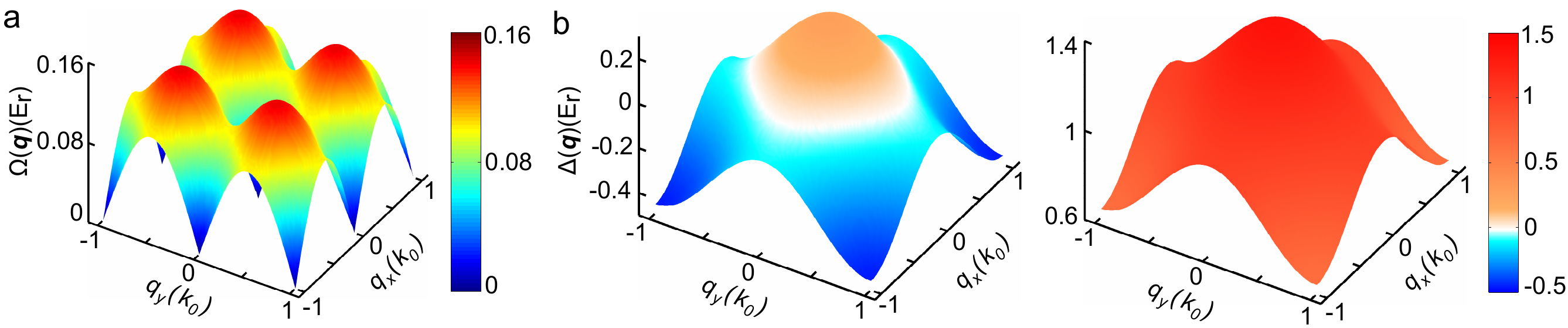}
\caption{\textbf{The momentum-dependent Raman coupling potential and local
gap.} \textbf{a}. The Raman coupling potential in the momentum space.
$\Omega(\bm{q})$ reaches maximum at $(\pm{0.5k_{0}})$
and minimum of $0$ at $\Gamma,M,X_{1}$ and $X_{2}$ points. \textbf{b}.
The local gap in the momentum space. $\Delta(\bm{q})$ has a \emph{ring}
with zero value in the left diagram but hasn't in the right diagram.
The parameters are $V_{0}=4E_{r}$, $\Omega_{0}=1E_{r}$, $\delta_{f}=0.2E_{r}$
(left) and $\delta_{f}=-2E_{r}$ (right), which corresponds to $t_{0}\approx0.09E_{r},t_{\text{SO}}\approx0.05E_{r}$.
The color bars represent the energy for $\Omega(\bm{q})$ and
$\Delta(\bm{q})$.}
\label{Randd}
\end{figure}

Our present system can be well described with a two-band
model. The optical lattice and the Raman coupling
lattice are all position-dependent and periodic, inducing spin-conserved
and spin-flipped tunneling between lattice sites, respectively. The Bloch Hamiltonian in the momentum
space can be obtained by~\cite{2DSOC2016}
\begin{equation}
\mathcal{H}(\bm{q})=\bm{h}(\bm{q})\cdot\bm{\sigma}=2t_{{\rm so}}\sin q_{y}\sigma_{x}+2t_{{\rm so}}\sin q_{x}\sigma_{y}+(-\delta_{f}/2+2t_{0}\cos q_{x}+2t_{0}\cos q_{y})\sigma_{z},\label{Hamiltonian}
\end{equation}
where $t_{0}$ denotes the spin-conserved tunneling, $t_{{\rm so}}$
represents the spin-flip tunneling term and $\bm{\sigma}$ is the
Pauli matrix. The hopping terms $t_{0}$ and $t_{{\rm so}}$ are determined
by the lattice depth and Raman coupling as
\begin{equation}
t_{0}=-\int{\rm d}\bm{r}\phi_{s}^{\vec{i}}(\bm{r})\left[\frac{\bm{p}^{2}}{2m}+V(\bm{r})\right]\phi_{s}^{\vec{i}+\vec{1}}(\bm{r}),
\end{equation}
and
\begin{equation}
t_{{\rm so}}=(-1)^{i}\int{\rm d}\bm{r}\phi_{s\uparrow}^{\vec{i}}(\bm{r})(M_{1}(\bm{r})+iM_{2}(\bm{r}))\phi_{s\downarrow}^{\vec{i}+\vec{1}}(\bm{r}),
\end{equation}
where $\phi_{s}$ is the Wannier function of s-band, and $\vec{i}$
represents lattice point. In the momentum space, the Raman coupling
strength $\Omega(\bm{q})$ and local gap $\Delta(\bm{q})$
can be written as $\Omega(\bm{q})=\sqrt{h_{x}^{2}(\bm{q})+h_{y}^{2}(\bm{q})}=2t_{{\rm so}}\sqrt{\sin^{2}q_{x}+\sin^{2}q_{y}}$,
and $\Delta(\bm{q})=h_{z}(\bm{q})=-\delta_{\text{f}}/2+2t_{0}\cos q_{x}+2t_{0}\cos q_{y}$,
which can be found that $\Omega(\bm{q})$ and $\Delta(\bm{q})$
are all momentum-dependent as shown in Fig.~\ref{Randd}. The local
gap $\Delta(\bm{q})$ can be regarded as the energy gap of the
Hamiltonian $\mathcal{H}(\bm{q})$ with Raman-induced hopping $t_{{\rm so}}\to0$,
which is also the gap of the normal bands.

Here we apply the above theory to our QAH system. The two $s$-bands of spin-up and spin-down intersect each other with a closed or opening gapless ring as showed in Fig.~\ref{bandtheory}(a-b), we denote the spin-up band as red, and spin-down
band as blue. When the Raman coupling is present, a gap opens
on the ring, and the spin polarization would be $0$ on this
ring. The total Chern number then can be determined as $\mathcal{C}=(-1)^{m}n$,
where the winding number $n=\frac{1}{2\pi}\text{\ensuremath{\oint}}_{\partial\mathcal{S}}\nabla_{\bm{q}}\theta_{\bm{q}}\cdot{\rm d}\bm{q}=-1$.
Here $\theta_{\bm{q}}={\rm arg}(M_{\bm{q}})$ and the integration
is always taken in a anticlockwise direction as shown in Fig.~\ref{bandtheory}(c-d).
Then the factor $m$ can be determined by the spectra of spin-up and spin-down normal bands. We choose
the spin-up band as a reference. For the case of Fig.~\ref{bandtheory}(a,c), the
Zeeman term $\delta_{\text{f}}>0$ and inside the ring the energy of the
spin-up band $E=E_{\uparrow}<0$. Hence, we have $m=1$ and the Chern number $\mathcal{C}=+1$.
For the case shown in Fig.~\ref{bandtheory}(b,d), the Zeeman term $\delta_{\text{f}}<0$
and inside the ring the energy of the spin-up band $E=E_{\uparrow}>0$. Hence,
we have $q=0$ and the Chern number $\mathcal{C}=-1$. In summary one has
\begin{eqnarray}\label{Chernnumber1}
{\cal C}= \left\{ \begin{array}{ll}
        +1, \ {\rm if \ the \ ring \ circles \ \Gamma \ point}, \\
             -1, \ {\mbox{if the ring circles M point}},\\
             0, \ {\mbox{no ring emerges}}.
        \end{array} \right.
\end{eqnarray}

While for present specific 2D QAH system, the Chern number takes relatively simple configurations with ${\cal C}=0,\pm1$, to determine the Chern number by 1D topological invariant (winding number) on rings can certainly be applied to generic case, as discussed in the previous section. In the present experiment, we extract the 1D topological invariant based on the ring patterns to determine the 2D Chern number, which can greatly simplify the
detection of topology of Chern bands. This strategy of measurement can be further generalized to other situations of characterizing a topological invariant of a higher dimensional system by a lower dimensional one, like the Chern-Simons invariant~\cite{Liu2017_1}.

\subsection*{C. Quench dynamics}

The quench dynamics of our experiment starts with the initial state
\begin{equation}
\left|\psi(\bm{q},t=0)\right\rangle =\left|\bm{q}\right\rangle \otimes\left|\uparrow\right\rangle .
\end{equation}
The state at time $t$ and quasi-momentum $\bm{q}$ of the quench
evolution can be expressed as ($\hbar\equiv1$)~\cite{ZWW2017}
\begin{align}
\left|\psi(\bm{q},t)\right\rangle =\begin{pmatrix}\alpha_{\bm{q}\uparrow}\\
\alpha_{\bm{q}\downarrow}
\end{pmatrix} & =e^{-\text{i}\mathcal{H}(\bm{q})t}\left|\psi(t=0)\right\rangle \nonumber \\
 & =\begin{pmatrix}\frac{h_{z}\left(-\text{i}\text{sin}\left(E_{\bm{q}}t\right)\right)}{E_{\bm{q}}}-\text{cos}\left(E_{\bm{q}}t\right)\\
\frac{\left(h_{x}+\text{i}h_{y}\right)\left(-\text{i}\text{sin}\left(E_{\bm{q}}t\right)\right)}{E_{\bm{q}}}
\end{pmatrix}\,,
\end{align}
where, $E_{\bm{q}}=\sqrt{h_{x}^{2}+h_{y}^{2}+h_{z}^{2}}=\sqrt{\Delta^{2}(\bm{q})+\Omega^{2}(\bm{q})}$
is the eigen-energy of our Hamiltonian. The expression at $E_{\bm{q}}\neq0$
for spin polarization observed in the experiment is
\begin{align}
P(\bm{q},t)= & |\alpha_{\bm{q}\uparrow}|^{2}-|\alpha_{\bm{q}\downarrow}|^{2}\,,\nonumber \\
= & \frac{h_{z}^{2}+\text{cos}\left(2E_{\bm{q}}t\right)\left(h_{x}^{2}+h_{y}^{2}\right)}{E_{\bm{q}}^{2}}\,,\nonumber \\
= & \frac{\Delta^{2}(\bm{q})+\text{cos}\left(2E_{\bm{q}}t\right)\Omega^{2}(\bm{q})}{\Delta^{2}(\bm{q})+\Omega^{2}(\bm{q})}\,,\label{spin_p}
\end{align}

where we can clearly see the time-dependent oscillation of spin polarization
with frequency $E_{\bm{q}}/\pi$ and amplitude $\Omega^{2}(\bm{q})/(\Delta^{2}(\bm{q})+\Omega^{2}(\bm{q}))$
from the expression in Eq.~\ref{spin_p}. The oscillating amplitudes
are all zero at the momentum points ($\Gamma,M,X_{1}$ and $X_{2}$)
for zero $\Omega(\bm{q})$. On the \emph{ring} of $\Delta(\bm{q})=0$,
the oscillating amplitude has maximum value of $1$.

\subsection*{D. The connection between ring and topology}

\begin{figure}[h]
\includegraphics[width=0.6\linewidth]{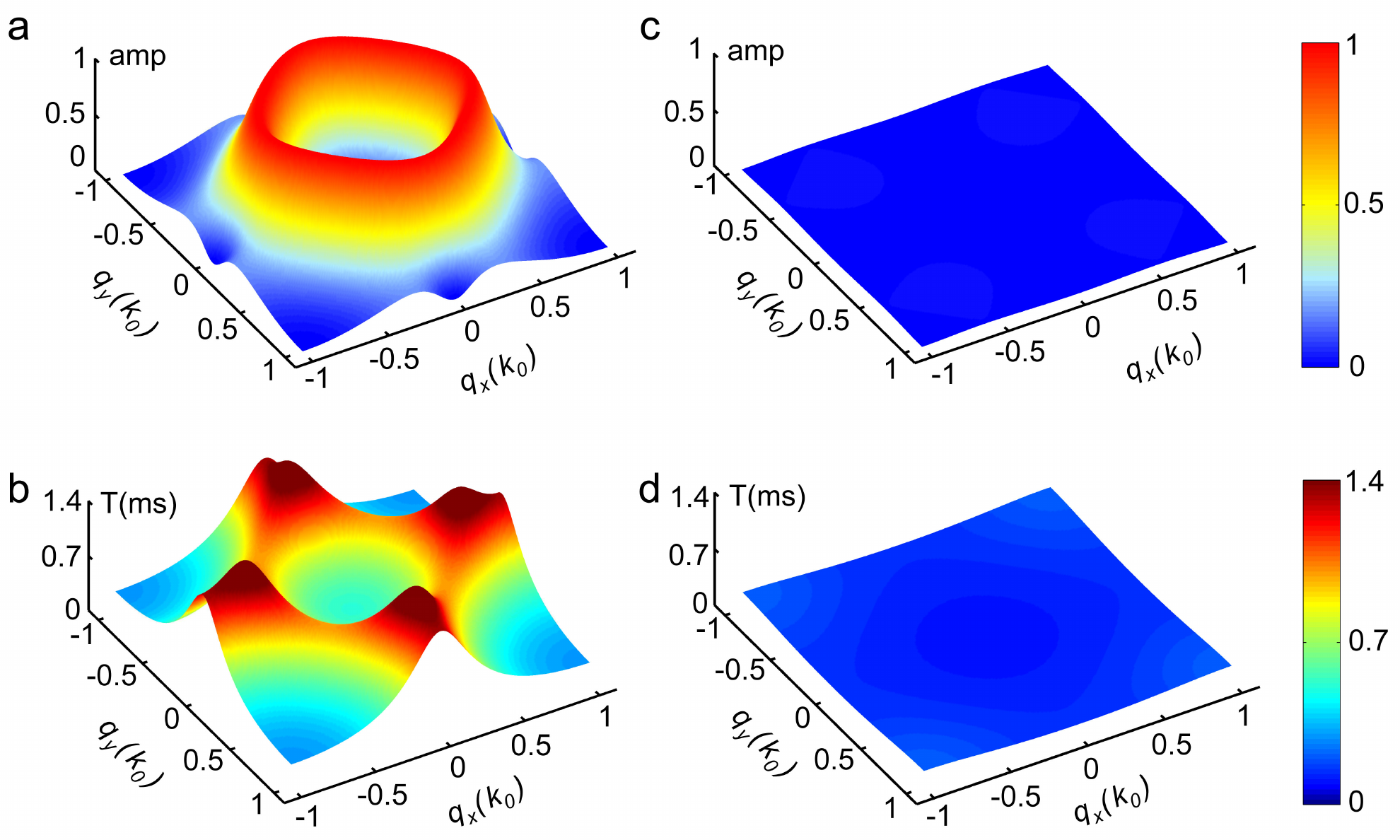}
\caption{\textbf{The oscillation amplitude and period distribution in FBZ.} \textbf{a} and \textbf{b} are the oscillation
amplitude and period distribution in FBZ while $C=1$ with $t_{0}=0.09E_{r},t_{\text{SO}}=0.05E_{r},\delta_{\text{f}}=0.2E_{r}$;
\textbf{c} and \textbf{d} are the oscillation amplitude and period
distribution while $C=0$ with $t_{0}=0.09E_{r},t_{\text{SO}}=0.05E_{r},\delta_{\text{f}}=-2E_{r}$.}
\label{A_P4}
\end{figure}

The topological non-trivial zone locates in the regime $-8t_{0}\leqslant\delta_{\text{f}}\leqslant8t_{0}$~\cite{Liu2013a},
where $\Delta(\bm{q})$ can be zero and form a ring. On this ring,
the amplitude of the spin polarization reaches the maximum value 1 according to Eq.~\ref{spin_p}
and the oscillating period is large as seen from \textbf{a} and \textbf{b}
in Fig.~\ref{A_P4}. As the initial state is spin up polarized in
the experiment, $P(\bm{q},t)$ can reach $-1$ in the time evolution.

The experimental observation of $P(\bm{q},t)=-1$ is shown in Fig.
2, which also forms a \emph{ring} structure. While $\delta_{\text{f}}<-8t_{0}$
or $\delta_{\text{f}}<8t_{0}$, the system is in the topological trivial
zone. The oscillating amplitude can not reach $+1$ and the oscillating
period is small as seen from \textbf{c} and \textbf{d} in Fig.~\ref{A_P4}.
There is no \emph{ring} formed in the spin dynamics.

The Chern number can be readily obtained from the ring pattern formed in the spin dynamics based on the formula~\eqref{Chernnumber1}. For the case $0<\delta_{\text{f}}<8t_{0}$, the dynamical ring circles the $\Gamma$ point [Fig.~\ref{A_P4}(a)]. Thus one has ${\cal C}=+1$. On the other hand, if $0>\delta_{\text{f}}>-8t_{0}$, one shall have that the dynamical ring circles $M$ point (not shown here), and then ${\cal C}=-1$. In experiment the dynamical ring pattern can be observed by spin- and momentum-resolved TOF imaging, as performed in this work.

The connection between local gap $\Delta(\bm{q})$ and the topology
of the system can also be understood with the spin index defined from the static spin polarization
distribution $S(\bm{q})=\left\langle \psi_{\text{f}}(\bm{q})\right|\sigma_{z}\left|\psi_{\text{f}}(\bm{q})\right\rangle $
where $\psi_{\text{f}}(\bm{q})$ is the eigenstate of $H_{\text{f}}$~\cite{Liu2013a}.
The momentum points forming the \emph{ring} structure with $\Delta(\bm{q})=0$
are exactly the locations with $S(\bm{q})=0$ on the equilibrium
lowest s-band of $H_{\text{f}}$. The topology of our system can be
detected by analyzing $S(\bm{q})$. Considering the highly symmetric
points $\Gamma$, $M$, $X_{1}$ and $X_{2}$ with $\Omega(\bm{q})=0$
in our topological non-trivial system, the sign of corresponding stable
spin polarizations $S(\bm{q})$ will have either three positive
and one negative signs or three negative and one positive signs. Specifically,
when the system $\mathcal{C}=-1$, $S(M)$ is negative and $S(\Gamma,X_{1},X_{2})$
are positive; when the system $\mathcal{C}=+1$, $S(\Gamma)$ is positive
and $S(M,X_{1},X_{2})$ are negative. Since the static spin polarization
$S(\bm{q})$ is smoothly distributed in FBZ, the locations of
\emph{ring} structure satisfying $S(\bm{q})=0$ separates
the regions of opposite static spin polarizations, and signifies the band crossing of the spin-up and spin-down bands. Accordingly, the Chern number ${\cal C}=+1$ and $-1$ when the ring circles the $\Gamma$ and $M$ points, respectively.

\subsection*{E. Topological phenomena in quantum quench as a  Continuous-time Quantum walk}

A two-dimensional continuous-time quantum walk (CTQW) analyzes the evolution of an particle moving through a 2D lattice. The Hamiltonian in Eq.~\ref{Hamiltonian} can be rewritten in lattice space as
\begin{equation}
\mathcal{H} =\sum_{xy}\left[
c^\dagger_{x+1y}\left(\text{i}t_\text{SO}\sigma_y+t_0\sigma_z\right)c_{xy}
+c^\dagger_{xy+1}\left(\text{i}t_\text{SO}\sigma_x+t_0\sigma_z\right)c_{xy}
+c^\dagger_{xy}\frac{\delta}{2}\sigma_zc_{xy}
\right]\,,
 \label{H_lattice}
\end{equation}
Where
\begin{equation}
c_{\bm{q}}=\frac{1}{\sqrt{N}}\sum_{{\bf r}_{xy}}c_{xy}e^{-\text{i}\bm{q}\cdot \bf r_{xy}}\,.
\end{equation}

The position is defined by ${\bf r}_{xy}=x{\bf a}_1+y{\bf a}_2$ with ${\bf a}_{1/2}$ as Bravais vectors of a 2D square lattice in x/y direction.
The first/second term in Eq.~\ref{H_lattice} describes the particle's hopping in x/y direction accompanied with spin operations, the third term is the onsite spin flip term.

\begin{figure}[h]
\includegraphics[width=0.75\linewidth]{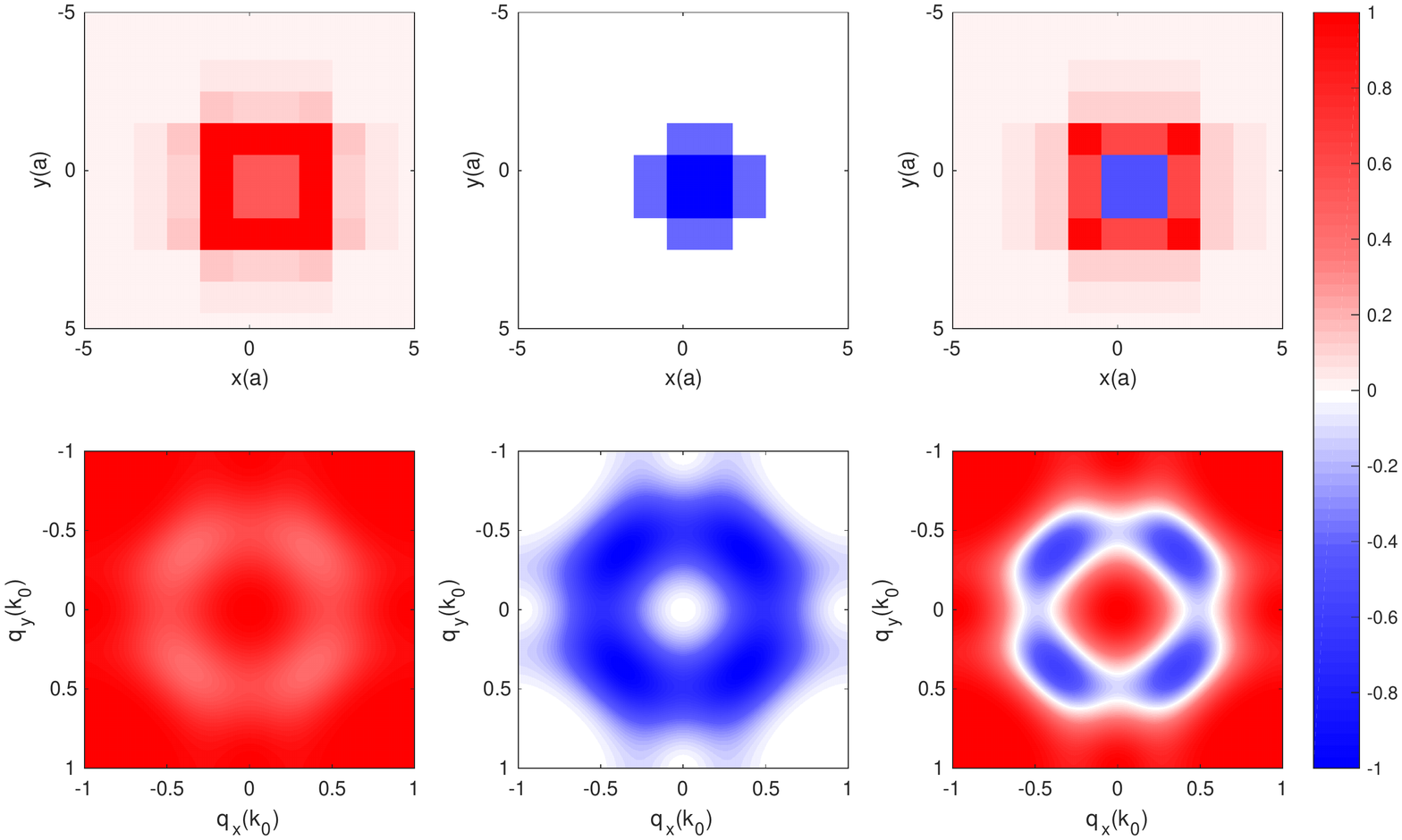}
\caption{\textbf{The CTQW dynamical patterns while $C=1$.}
First row are the CTQW evolution patterns in lattice space  (from left to right: spin up component, spin down component, total)  while $C=1$ with $t_0=0.09E_{r}, t_{\text{SO}}=0.05E_{r}, \delta=0.4E_{r}$; Second row are the CTQW patterns in momentum space  (from left to right: spin up component, spin down component, spin polarisation) with the same system as in the first row.}
\label{C1}
\end{figure}

The CTQW dynamics of a particle initially localized in the centre of the lattice with spin up internal state can detect the system's topological phase, where a characteristic central feature in the density profile of the particle's probability distribution on lattice displays the topological nontrivial phases~\cite{ZWW2017}.
The appearance of central feature in the particle's density profile with a topological non-trivial Hamiltonian is mainly contributed from the momentum components at minimum bandgap.
Qualitative changes to the central features of particle's density profile, including its size, disappearance and reappearance under system parameter changes, are closely related with the topological phase of the system.
%
%

The quench dynamics in our experiment is an analogous to a CTQW, where the state before quench is analogous to the particle's initial state in CTQW and the final Hamiltonian of quench is analogous to particle's system Hamiltonian in CTQW. The broad momentum distribution of atoms before quench dynamics is analogous the particle's localized initial state in lattice space of CTQW.
Our quench observations in momentum space directly reveal the dominated momentum contribution to the central features of density profile in lattice space.
The \emph{ring} structure we observed in experiment is the corresponding minimum bandgap locations in momentum space and can be obtained through the Fourier transformation of the particle's central peak components in lattice space which is a Bessel function as shown in~\cite{ZWW2017}.
To demonstrate the analogous of the quench dynamics and CTQW,
we consider one particle walking in our 2D SO coupling system with the lattice constant $a=\lambda/2$, lattice depth $V_{0}=4E_\text{r}$, the Raman coupling strength $\Omega_{0}=1E_\text{r}$. The corresponding spin-conserved tunneling rate $t_{0}$ is about $0.09E_\text{r}$ and the spin-flipped tunneling rate $t_\text{SO}$ is about $0.05E_\text{r}$.
In Fig.~\ref{C1}, we show the CTQW dynamical patterns in both lattice space and momentum space with topological non-trivial Hamiltonians having $\mathcal{C}=+1$. The first row shows the particle's spin distributions in lattice space (from left to right: spin up, spin down, spin polarization) and the second row shows the particle's spin distributions in momentum space  (from left to right: spin up, spin down, spin polarization).
These plots mimic the particle's short time evolutions (evolution time $t/\hbar$=5), which corresponds to the evolution time of $350\mu s$. There are centre features in the spin down components of particle's density profile in lattice space and the \emph{ring} patterns in momentum space.
The appearance of both momentum \emph{ring} structure in our experimental quantum quench dynamics and the \emph{ring} structure in CTQW's momentum patterns comes from the minimum bandgap locations which is also responsible for the appearance of central feature in particle's density profile in lattice space.
The quantum quench dynamics and CTQW both are rooted in the topology of the system's band structure and ensure the discovery of the system's  topological phase.

\subsection*{F. Mapping the phase diagram of band topology based on the size of ring}

With the interpretation of the connection between ring and topology being established, we can further determine the full phase diagram of the band topology by examining the evolution of ring size in the spin dynamics.
We find that with the detuning $\delta_{\text{f}}$ increasing (decreasing)
\begin{figure}[h]
\includegraphics[width=0.5\linewidth]{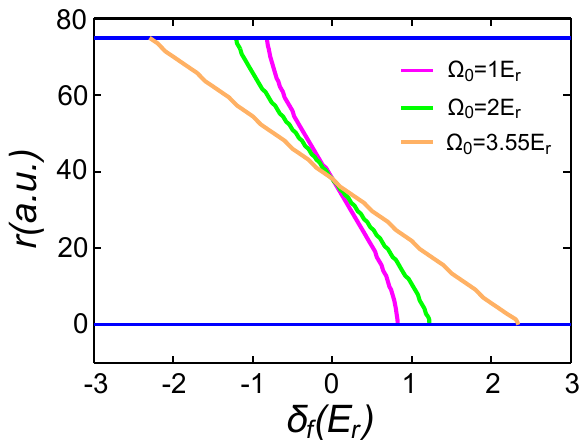}
\caption{\textbf{The numerical simulation of the }\textbf{\emph{ring}}\textbf{ sizes versus
$\delta_{\text{f}}$.} The Raman coupling strengths $\Omega_{0}$ are $1E_{r}$
(magenta line), $2E_{r}$ (green line) and $3.55E_{r}$ (yellow line).
The blue lines located at the top and bottom are the maximal and minimal
\emph{ring} sizes respectively.}
\label{size}
\end{figure}
, the ring would shrink and disappear at $\Gamma$ ($M$) point right
after $|\delta_{\text{f}}|>\delta_{c}$. Hence, we can determine the topological
phase boundaries by observing the evolution of the ring pattern versus
the final two-photon detuning $\delta_{\text{f}}$. Here, the \emph{ring}
size is defined as the distance from the $\Gamma$ point to the \emph{ring} along the diagonal direction as shown in Fig.~\ref{bandtheory}.

The numerical simulation of the \emph{ring} size versus the detuning $\delta_{\text{f}}$ shows a nonlinear relation between them as shown in Fig.~\ref{size}. We use these curves in a finite-terms polynomial fitting to obtain the value of the detuning $\delta_\text{f}=\delta_\text{c}$ on the topological boundaries.


\section*{III. Data analysis and additional datasets}

\subsection*{A. Topological phase probed with \emph{ring} structure, extracting the phase diagram}

\subsubsection{Patterns of spin polarisation in short-time evolution}

The time evolution of the spin polarisation after the quantum quench is only weakly damped and for short times is quasi-unitary.
\begin{figure}[h]
\includegraphics[width=0.9\linewidth]{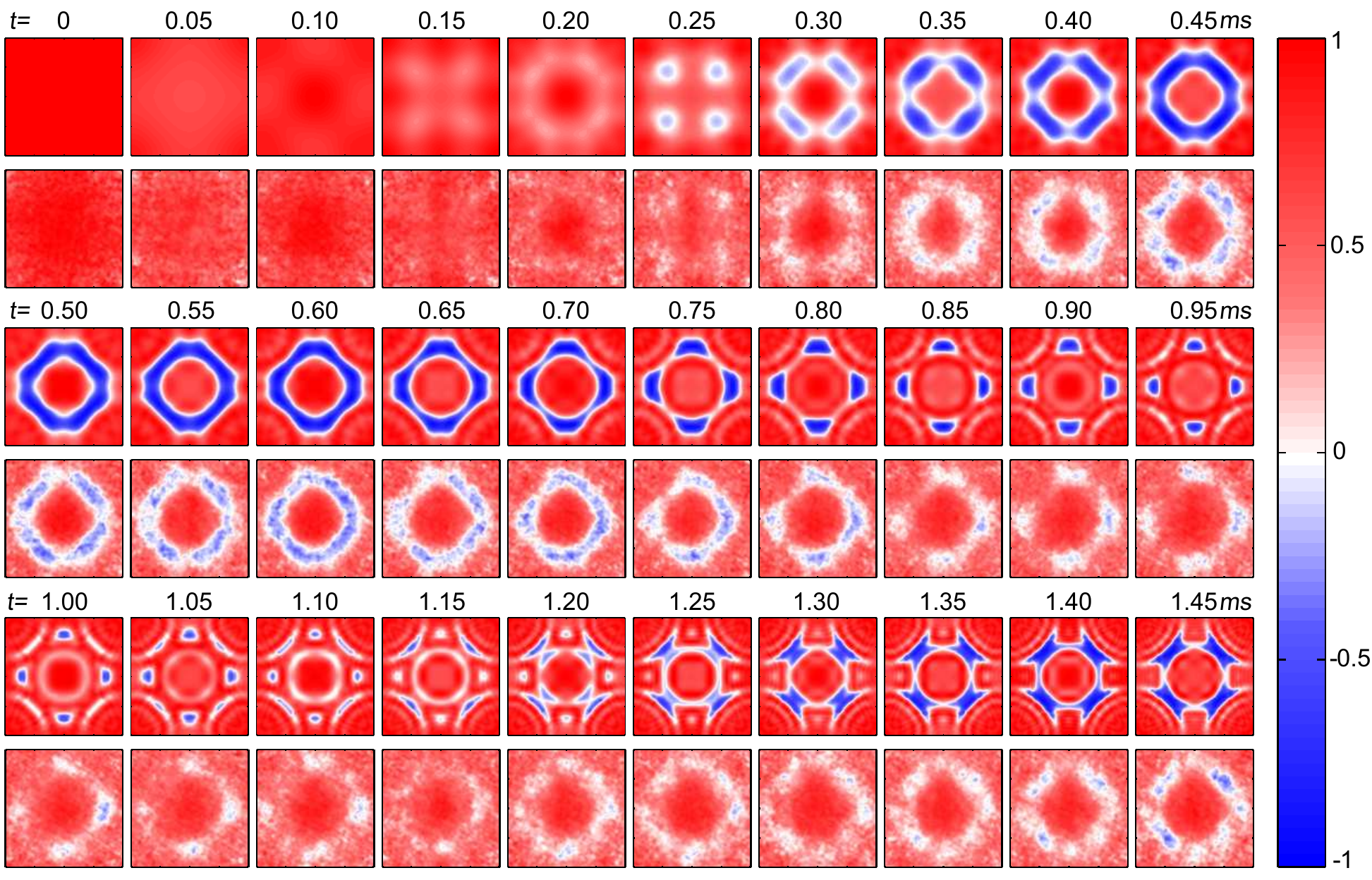}
\caption{\textbf{Patterns of spin polarisation in the FBZ with short time evolution.} The measurement is under parameters $V_0=4E_{\text{r}}$, $\Omega_0=1E_{\text{r}}$ and $\delta_f=0.2E_{\text{r}}$. The first, third and fifth rows are the numerical simulations and the second, forth and sixth rows the experimental data . The time is from $0$ to $1.45ms$ with time step of $50\mu s$.}
\label{pattern_short}
\end{figure}
 Its pattern effectively reveals the appearance of \emph{ring} structure. A typical set of data for $V_{0}=4E_{r}, \Omega_{0}=1E_{r}, \delta_{f}=0.2E_{r}$ is shown in Fig.~\ref{pattern_short}. The first, third and fifth rows are the numerical simulations. The second, forth and sixth rows are the experimental measurement. The initial spin polarization at $t=0$ is fully spin up over the FBZ, then starts oscillating after quench. Around the momentum with largest $\Omega(\bm{q})$ and zero $\Delta(\bm{q})$, the spin polarization can flip to large negative value and shows \emph{ring} structure both in theory and experiment. At $t=0.25ms$, the spin polarization firstly oscillates to negative value where $\Omega(\bm{q})$ is largest. The other momentum on the \emph{ring} with smaller $\Omega(\bm{q})$ need longer time to reach negative spin polarization, which shows a trajectory of the negative spin polarization moving from one momentum place to other momentum place at about $t=1ms$ on the \emph{ring}. The negative spin polarization then moves back at about $t=1.45ms$ and shows an oscillation versus time. The trajectory of negative spin polarization circles the positive spin polarization at the $\Gamma$ point and gives the Chern number $\mathcal {C}=+1$.

\subsubsection{Determining topological phase boundary under higher $\Omega_0$}

\begin{figure}[h]
\includegraphics[width=0.9\linewidth]{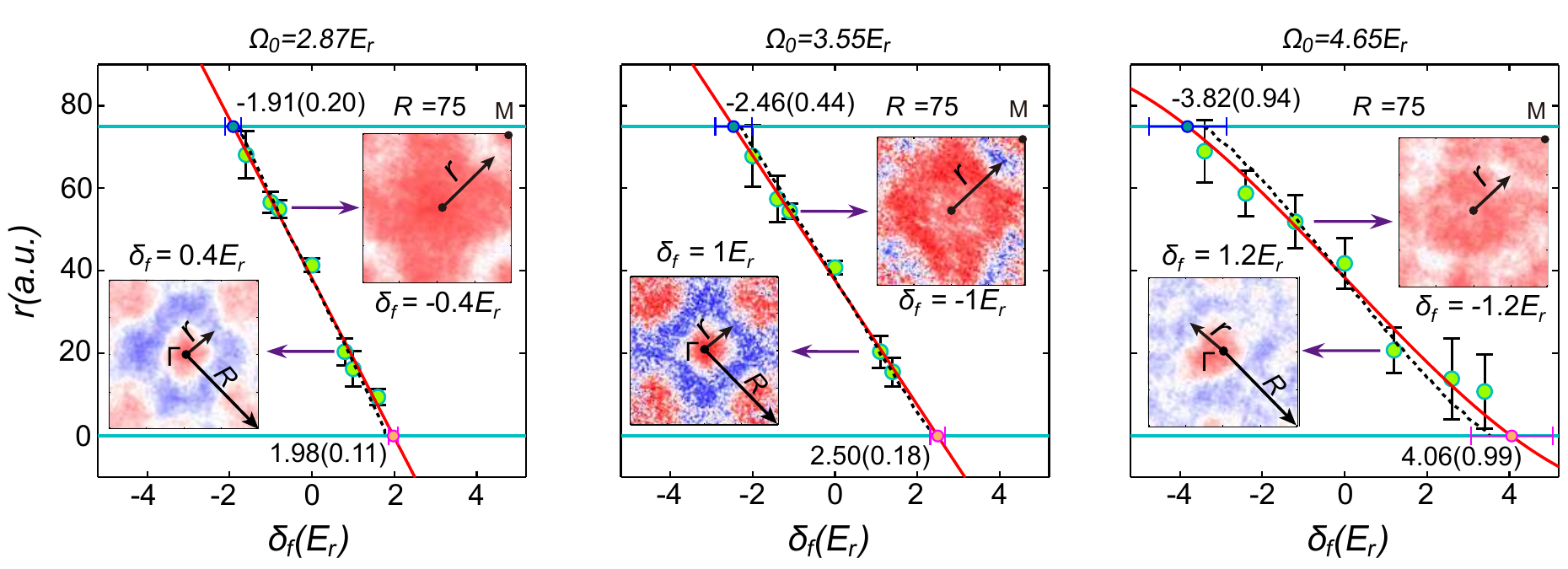}
\caption{\textbf{Determining topological phase boundary under higher $\Omega_0$}.
The Raman coupling strengths from left to right are $\Omega_0=2.87E_{r}, 3.55E_{r}$ and $4.65E_{r}$, respectively. The green dots are the sizes $r$ of the \emph{ring}. The red line is the polynomial fitting up to
3rd order. The inserts are the typical patterns at different detunings. The blue lines represent the minimal and maximal value of the \emph{ring} and have two cross points (blue dot and red dot with error barsp) with the red line. The detunings at blue and red dots determine the topological phase boundary.}
\label{phase}
\end{figure}

The appearance of the \emph{ring} structure shows that the system belongs to a topological nontrivial phase. The size of the \emph{ring} can determine the phase boundary precisely by fitting the size at different $\delta_{\text f}$. We have five data sets to map the topological phase boundary. Two fitting data with low Raman coupling strength are shown in Fig. 3. Here we show the other three with high Raman coupling strengths $\Omega_0=2.87E_{r}, 3.55E_{r}$ and $4.65E_{r}$, respectively (Fig.~\ref{phase}). The phase boundaries are at $\delta_{\text f}=-1.91\pm 0.20$ and $1.98\pm 0.11$ under $\Omega_{0}=2.87E_{r}$, $\delta_{\text f}=-2.46\pm 0.44$ and $2.50\pm 0.18$ under $\Omega_{0}=3.55E_{r}$, $\delta_{\text f}=-3.82\pm 0.94$ and $4.06\pm 0.99$ under $\Omega_{0}=4.65E_{r}$.

\subsection*{B. Analysing the long term evolutions}

In the experiment, the long term evolution after the quantum quench shows a relaxation process. The relaxation can have many possible sources like atom loss, internal relaxation, pure dephasing~\cite{Matsuo2009} or the contact interaction between atoms~\cite{Decaymodel}. In our system, the atoms loss is negligible in the time scale of miliseconds~\cite{2DSOC2017}. The internal relaxation represents the atoms decay from high bands to low bands after the quench and leads to an exponential decay. The pure dephasing in the resonant oscillation comes from noise in the experiment like magnetic noise and leads to an exponentially damped oscillation. The contact interaction will add an extra term in the single particle Hamiltonian and cause system decay from non-equilibrium state towards equilibrium state~\cite{Decaymodel}.

 At a momentum point $\bm{q}_{1}$ close to the \emph{ring}, the Raman coupling strength $\Omega(\bm{q})$ is large and the local gap $\Delta(\bm{q})\approx0$, the Raman-Rabi oscillation is nearly resonant. The spin-up and spin-down bands cross here with Raman coupling strength $M_{0}\rightarrow 0$ as shown in Fig.~\ref{bandtheory}. The eigenenergy of both spin here is near degenerate, leads to the internal relaxation negligible. The decay mainly comes from the pure dephasing and causes a damped oscillation. The contact interaction leads the atoms to the equilibrium state finally. The spin polarization is fitted in the function~\cite{Matsuo2009,Decaymodel} of
\begin{equation}
F(t)=Ae^{-\frac{t}{t_1}}\cos{(ft+\phi)}+Be^{-\frac{t}{t_{2}}}+Ce^{-(\frac{t}{t_{3}})^2}+D.
\end{equation}
For the momenta close to $X_1$ and $X_{2}$, $\Omega(\bm{q})$ is near zero and $\Delta(\bm{q})\approx\delta_{\text{f}}$. There is no dephasing effect. The decay is dominated by the internal relaxation and the contact interaction. The fitting function~\cite{Matsuo2009} is
\begin{equation}
F(t)=Be^{-\frac{t}{t_{2}}}+D.
\end{equation}
For the momenta at the $\Gamma$ and the $M$ points, $\Omega(\bm{q})=0$ and $\Delta(\bm{q})$ is large, the eigenenergy of the lowest s-band are local minima. Neither the internal relaxation nor the dephasing is obvious. The dominant process comes from the atoms interaction. The fitting function~\cite{Decaymodel} has the form of
\begin{equation}
F(t)=Ce^{-(\frac{t}{t_{3}})^2}+D.
\end{equation}

\begin{table}[h]
\caption{Fitting parameters of Fig. 5}
\begin{tabular}{|p{0.5cm}<{\centering}|*{9}{p{1.7cm}<{\centering}|}}
\hline
  & $A$ & $t_{1}/ms$ & $f/Hz$ & $\phi$ & $B$ & $t_{2}/ms$ & $C$ & $t_{3}/ms$ & $D$ \\
\hline
$\Gamma$ & $-$ & $-$ & $-$ & $-$ & $-$ & $-$ & $0.38(0.03)$ & $4.04(0.22)$ & $0.48(0.01)$ \\
\hline
$M$      & $-$ & $-$ & $-$ & $-$ & $-$ & $-$ & $1.05(0.04)$ & $5.38(0.29)$ & $-0.48(0.02)$ \\
\hline
$X_{1}$  & $-$ & $-$ & $-$ & $-$ & $0.72(0.06)$ & $2.34(0.19)$ & $-$ & $-$ & $-0.15(0.01)$ \\
\hline
$\bm{q}_{1}$ & $0.67(0.13)$ & $1.27(0.22)$ & $1012(23)$ & $0.00(0.00)$ & $0.56(0.10)$ & $2.01(1.20)$ & $-0.36(0.1)$ & $7.67(0.54)$ & $(0.04)(0.03)$ \\
\hline
\end{tabular}\\
\end{table}

The values of the fitting parameters in Fig. 5 of main text are shown in the Table I, where the timescale of atoms interaction is about $5.3$ms, the the time scale of external decay is about $2.8$ms and the timescale of dephasing is about $1.6$ms.

\begin{figure}[h]
\includegraphics[width=0.7\linewidth]{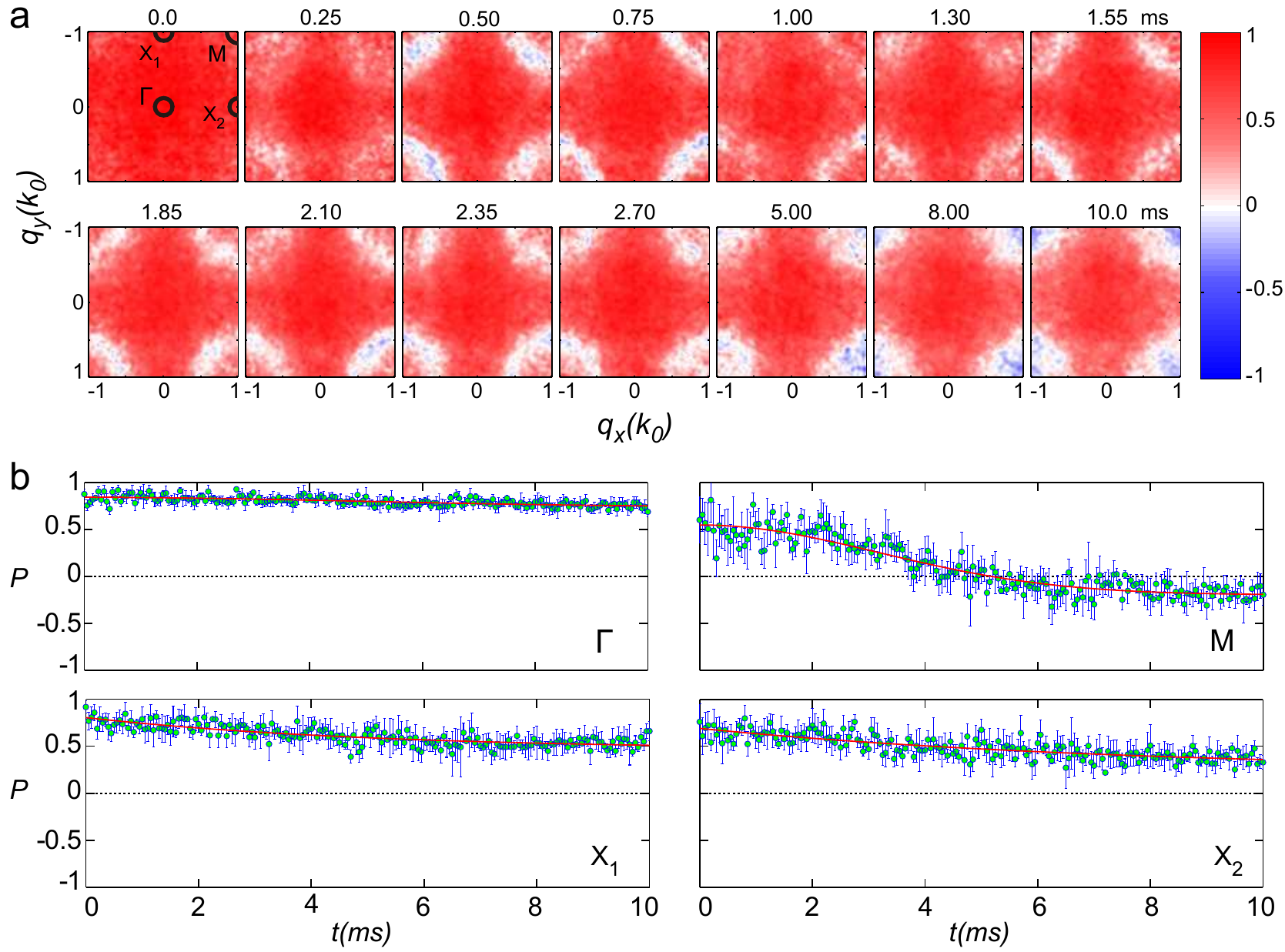}
\caption{\textbf{Long term evolution to equilibrium state.} \textbf a. The patterns of the spin polarisation in FBZ under parameters $V_0=4E_{\text{r}}$, $\Omega_0=1E_{\text{r}}$ and $\delta_{\text{f}}=-0.4E_{\text{r}}$. \textbf b. The long term evolution of the spin polarization at $\Gamma, M, X_{1}$ and $X_{2}$ points. The blue dots with error bars are the time evolution of the spin polarisation in the experiment. 
}
\label{long_evolution}
\end{figure}

With a long time evolution, the atoms reach the thermal equilibrium state of the final Hamiltonian which can be used to
detect the topological phase of the system~\cite{2DSOC2016,2DSOC2017}. The Chern number $\mathcal{C}$ of the system  can be detected by the signs of the spin polarisations at $\Gamma$, $M$, $X_1$ and $X_2$ points of the thermal equilibrium state.

The long term evolution of the spin polarisation under a system with $V_0=4E_\text{r}$, $\Omega_0=1E_\text{r}$ and $\delta_\text{f}=-0.4E_\text{r}$ is shown in Fig.~\ref{long_evolution}.
The atoms in the initial state with temperature about $100\text{nK}$ are heated to about $130\text{nK}$ at $t=10ms$. The temperature is not too high and the final thermal equilibrium states revealing the topology of the lowest band properly is accessible.

The spin polarisation at $\Gamma$, $M$, $X_1$ and $X_2$ points during a long quench dynamics are shown in  Fig.~\ref{long_evolution}\textbf{b}. The data is fitted and gives the final values of the spin polarization at $\Gamma$, $M$, $X_1$ and $X_2$ to $0.73\pm0.02, -0.2\pm0.02, 0.44\pm0.07$ and $0.23\pm0.09$,
which agree with the theoretical values ($0.69, -0.23, 0.3$ and $0.3$) of thermal equilibrium state at the temperature $100\text{nK}$. The Chern number $\mathcal C=-1$ is obtained from the signs of the spin polarisations at the four momentum points.

\end{document}